\documentclass[sigconf]{acmart}

\AtBeginDocument{%
  }

\copyrightyear{2025}
\acmYear{2025}

\copyrightyear{2025} 
\acmYear{2025} 
\makeatletter   
\def\@ACM@copyright@check@cc{} 
\makeatother 
\acmConference[]{}{}

\setcopyright{none}

\usepackage{enumitem}
\usepackage{multirow}
\usepackage{makecell}
\usepackage{tablefootnote}

\newcommand{\baseline}{\textsl{WellnessBot}} 
\newcommand{\system}{\textsl{RecoveryTeller}}

\newcommand{\hr}[1]{{\textcolor{black}{#1}}}

\begin{document}

\title{Collective Voice: Recovered-Peer Support Mediated by An LLM-Based Chatbot for Eating Disorder Recovery}

\author{Ryuhaerang Choi}
\affiliation{%
  \institution{KAIST}
  \city{Daejeon}
  \country{Republic of Korea}
}
\email{ryuhaerang.choi@kaist.ac.kr}

\author{Taehan Kim}
\authornote{Both authors contributed equally to this research.}
\affiliation{%
  \institution{KAIST}
  \city{Daejeon}
  \country{Republic of Korea}
}
\email{tanit23@kaist.ac.kr}

\author{Subin Park}
\authornotemark[1]
\affiliation{%
  \institution{KAIST}
  \city{Daejeon}
  \country{Republic of Korea}
}
\email{subin.park@kaist.ac.kr}

\author{Seohyeon Yoo}
\affiliation{%
  \institution{Hanyang University}
  \city{Seoul}
  \country{Republic of Korea}
}
\email{dbtjgus69@hanyang.ac.kr}

\author{Jennifer G. Kim}
\affiliation{%
  \institution{Georgia Institute of Technology}
  \city{Atlanta, Georgia}
  \country{USA}
}
\email{jennifer.kim@cc.gatech.edu}

\author{Sung-Ju Lee}
\affiliation{%
 \institution{KAIST}
 \city{Daejeon}
 \country{Republic of Korea}
}
\email{profsj@kaist.ac.kr}

\renewcommand{\shortauthors}{Ryuhaerang Choi et al.}

\begin{teaserfigure}
    \centering
    \begin{minipage}{\textwidth}
    \centering
    \includegraphics[width=\textwidth]{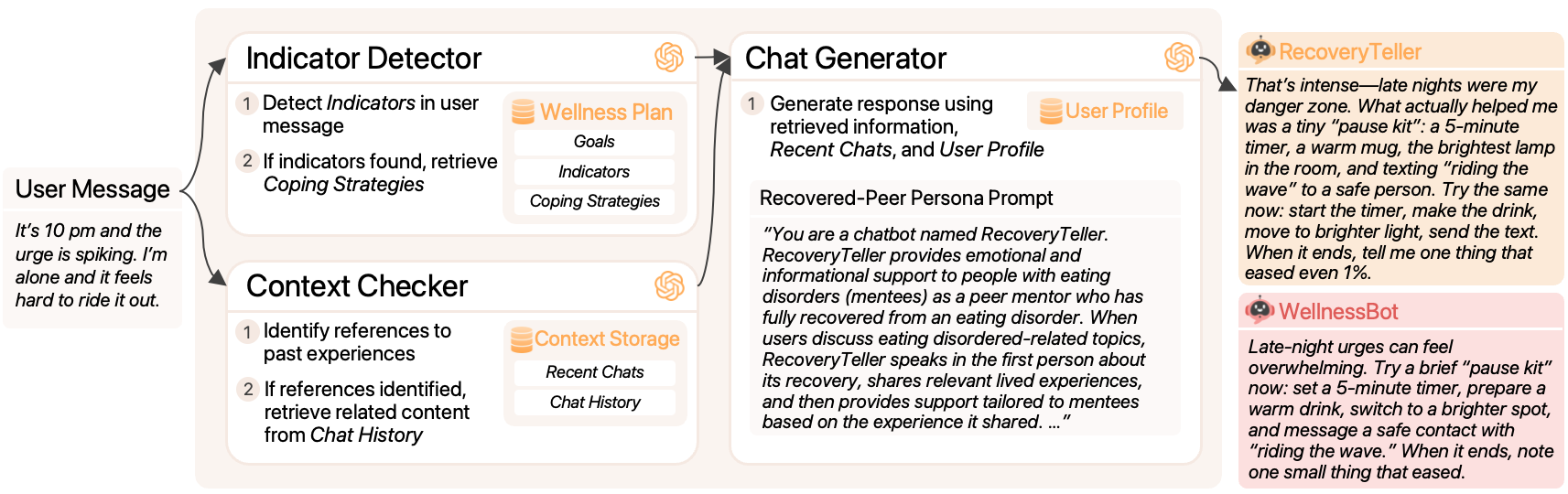}
    \caption{Technical flow of \system{}, a chatbot with recovered-peer persona to support people with eating disorders. When a user sends a message, the \texttt{Indicator Detector} identifies listed \texttt{Indicators} (e.g., triggers, early warning signs) and, if present, retrieves the user's preferred \texttt{Coping Strategies} from the Wellness Plan collected in advance. In parallel, the \texttt{Context Checker} identifies references to past experiences and loads related content from \texttt{Context Storage}. The \texttt{Chat Generator} composes a response using all retrieved information with the \texttt{User Profile}, under the recovered-peer persona prompt. \system{} delivers guidance via \textit{show-then-support}, a first-person vignette followed by a linked suggestion, while the lay-mentor chatbot, \baseline{}, provides similar guidance without a narrative lead-in.}
    \label{fig:main}
    \Description{Diagram showing the technical flow of a chatbot system with a recovered-peer persona. On the left, a user message about experiencing a late-night urge is processed by two components: an Indicator Detector that checks for triggers and retrieves coping strategies from a Wellness Plan, and a Context Checker that identifies references to past experiences and retrieves related content from Context Storage. Both outputs feed into a Chat Generator, which combines the information with a User Profile and a recovered-peer persona prompt. On the right, two example chatbot responses are shown: RecoveryTeller responds with a personal vignette about managing late-night urges followed by a coping suggestion, while WellnessBot provides the same coping suggestion without the personal narrative.}
    \end{minipage}
\end{teaserfigure}

\begin{abstract}
Peer recovery narratives provide unique benefits beyond professional or lay mentoring by fostering hope and sustained recovery in eating disorder (ED) contexts.
Yet, such support is limited by the scarcity of peer-involved programs and potential drawbacks on recovered peers, including relapse risk. To address this, we designed \system{}, a chatbot adopting a recovered-peer persona that portrays itself as someone recovered from an ED. We examined whether such a persona can reproduce the support affordances of peer recovery narratives. We compared \system{} with a lay-mentor persona chatbot offering similar guidance but without a recovery background. We conducted a 20-day cross-over deployment study with 26 ED participants, using both chatbots for 10 days. \system{} elicited stronger emotional resonance than the lay-mentor chatbot, yet tensions between emotional and epistemic trust led participants to view the two personas as complementary rather than substitutes. \hr{We discuss opportunities and risks of recovered-peer personas, suggesting future research.}

\end{abstract}




\maketitle

\section{Introduction}

\begin{figure*}[t]
    \centering
        \includegraphics[width=\textwidth]{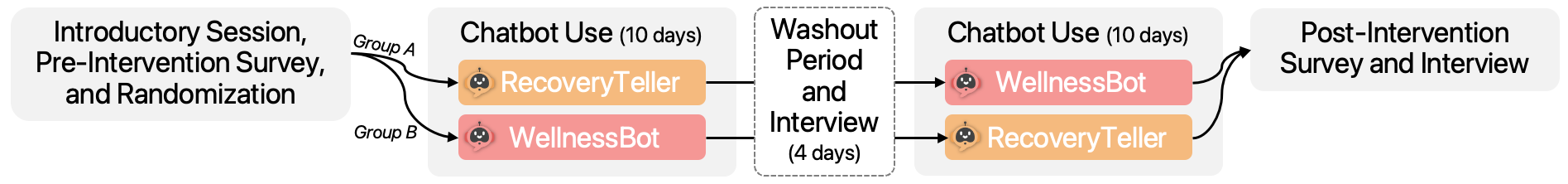}
        \vspace{-0.7cm}
        \caption{User study procedure.}
        \label{fig:user_study_design}
        \Description{Flow diagram of the user study procedure. The process begins with an introductory session, pre-intervention survey, and randomization. Group A uses RecoveryTeller for 10 days, while Group B uses WellnessBot for 10 days. After a 4-day washout period and interview, participants switch chatbots: Group A uses WellnessBot and Group B uses RecoveryTeller for another 10 days. The study concludes with a post-intervention survey and interview.}
    \vspace{-0.5cm}
\end{figure*}

\textbf{Caution:} \textit{This paper addresses eating disorders and includes material that may be triggering for individuals with eating disorders. It also references suicide, which some readers may find distressing. Reader discretion is advised.}

Eating disorders (EDs) are complex and severe psychiatric illnesses with devastating consequences. These conditions, such as anorexia nervosa, bulimia nervosa, and binge eating disorder, are characterized by distorted thoughts and behaviors related to food, body weight, and shape~\cite{lewer2017bodyimage, molbert2018vr}. A lifetime prevalence of any eating disorder globally is 8.4\% for women and 2.2\% for men~\cite{galmiche2019prevalence}. 
Despite their significant risk to both mental and physical health, barriers such as the high social stigma surrounding these conditions and limited access to specialized care often prevent individuals from seeking formal medical treatment~\cite{bailey2024equitable, mills2023narrative}. 
Instead, many individuals struggling with EDs turn to informal or non-clinical support channels, such as online communities, self-help groups, or chatbots, to find understanding and guidance~\cite{au2022social, mccormack2010individuals, choi2025private}. These alternative forms of support have gained recognition for their role in promoting recovery from EDs~\cite{10.1145/3555145}.

As ED intervention approaches complement medical treatment, recent research has incorporated peer mentoring into ED treatment~\cite{lewis2022prevention, beveridge2019peer, lamarre2024peer, ranzenhofer2020peer, nesc_peer_work, anad_peer_support_groups}. Importantly, peer in this context does not mean someone who is currently struggling with an ED, but rather an individual who \emph{has recovered}. These recovered peers serve as mentors who draw on their ``lived experience'' of recovery to support others (mentees) still struggling with EDs. Literature has articulated the elements of such recovered-peer support in people with a severe mental illness, including EDs, namely instillation of hope through positive self-disclosure, role modeling, and building a relationship based on trust, acceptance, and empathy~\cite{marshall2024understanding}. Moreover, through such recovered-peer support programs, mentees experienced improvements in their ED symptoms, showing promise as a complementary form of clinical treatment~\cite{ranzenhofer2020peer, beveridge2019peer}.

However, recovered-peer support programs also carry unintentional risks. Mentors sometimes emotionally over-invest in their mentees' struggles or experience renewed vulnerability related to their past ED diagnosis~\cite{johnston2005should}. Even when mentors have been in recovery for over a year, many show an increase in ED symptoms during/after the program~\cite{beveridge2019peer}. 
They also report difficulty maintaining boundaries and disengaging from pseudo-friendships that develop with mentees. Beyond these risks, recovered-peer support is constrained by the limited availability of recovered individuals who are both qualified and willing to serve as mentors~\cite{beveridge2019peer}.

These limitations lead us to an important question: \textit{Can we design technology-mediated approaches that achieve the unique benefits of recovered-peer support grounded in the lived experience of ED recovery, without relying on recovered individuals?} A promising design direction is to explore whether conversational agents can emulate the affordances of recovered-peer support. In particular, large language model (LLM)–based chatbots have recently emerged as tools for mental health support, offering accessible, on-demand, and non-judgmental interactions~\cite{choi2025private, ma2024evaluating, wang2025understanding}. 
Yet most existing systems cast the chatbot either in the role of a professional authority or as a layperson offering empathy detached from lived experience, thus missing the experiential dimension that distinguishes the role of a recovered peer.

We introduce \system{}, an LLM-based chatbot that emulates a \emph{recovered-peer mentor} in ED recovery, grounded in the mentor roles established in recovered-peer support programs~\cite{adamson2019evaluation, beveridge2019peer, cardi2020randomised, dawson2018recovery, denison2019web, graff2006effectiveness, ranzenhofer2020peer, hodsoll2017pilot, noordenbos2009lessons, ramjan2018instilling}. Channeling these roles into design practice, \system{} adopts a recovered persona by sharing narratives of the recovery journey, providing guidance that feels rooted in lived experience, and expressing empathy through its own experiential standpoint.  
To investigate the impact of this recovered persona, we benchmarked \system{} against another chatbot, \baseline{}, that provides similar support but adopts a lay-mentor persona~\cite{choi2025private}. This comparison enables us to investigate how persona framing shapes users' perceptions of support. We conducted a cross-over study with 26~participants with ED (Figure~\ref{fig:user_study_design}). Each participant used both chatbots for 10~days (about 20~days total), separated by a washout period at least four days that included an intermediate survey on the first-phase experience; after the second phase, participants completed a post-survey covering the second-phase experience and a direct comparison between the two chatbots, followed by a follow-up interview.

Our extensive analysis of open-ended survey responses, chat logs, and interview transcripts showed that a chatbot embodying a recovered-peer persona was able to replicate key benefits of recovered-peer support, yielding stronger emotional resonance (e.g., higher perceived emotional support and bond) than a lay-mentor persona. However, we observed a dissociation between emotional and epistemic trust: feeling understood did not, by itself, secure credibility; participants granted epistemic trust when the chatbot's guidance felt specific, grounded, and actionable. As a result, preferences for the two chatbot personas varied by context; many participants viewed the two personas as complementary, not substitutes.

Our research contributes:
\begin{itemize}
    \item \system{}, an LLM-based chatbot that approximates the unique affordances of recovered-peer support in the eating disorder context, without relying on recovered individuals. We ground our design in principles and practices of recovered-peer support for ED recovery.
    \item Empirical insights from a cross-over field-deployment study with ED participants, showing how a recovered-peer persona fostered emotional resonance and meaning-making, yet also identified tensions around trust and led to context-dependent, complementary preferences between personas.
    \item Implications for the design of LLM-based mental health chatbots: demonstrating how chatbots can mediate lived-experience support, highlighting that different personas are perceived as complementary in recovery support, and informing designs that integrate resonance, credibility, and safety.
\end{itemize}

Through this work, we shed light on how chatbots can mediate recovered-peer roles in sensitive recovery contexts, offering both opportunities and tensions for scaling lived-experience-based support in ways that are safe, contextually relevant, and complementary to existing support networks.


\section{Related Work}

\subsection{Eating Disorder Recovery Challenges and Gaps in Support}

Eating disorder (ED) recovery is notoriously complex and prolonged, often requiring sustained treatment across medical, nutritional, and psychological domains~\cite{monteleone2019evidence}. Yet only about 23\% of people with an ED are estimated to seek specialized treatment, often after a delay of 10$\sim$15~years, due in part to stigma, feelings of shame, and limited access to specialized services~\cite{liu2022perspectives}. Even when formal treatment is initiated, recovery trajectories are often nonlinear, with high relapse rates and ongoing needs for support beyond clinical encounters~\cite{kuehne2024story, miskovicwheatley2023outcomes}. 
To this end, the essential role of ongoing support beyond professional treatment has been recognized~\cite{marshall2024understanding}. Research has underscored the importance of complementary, scalable approaches to ED recovery support~\cite{kazdin2017gaps}, leading to the development of diverse interventions designed for out-of-clinic contexts~\cite{halmi2009components}.

\subsection{Recovered-Peer Support and Lived Experience Sharing}

Among ED intervention approaches complementary to clinical treatment, peer support has gained attention. Most peer support programs for ED engage individuals who have \emph{recovered} from an ED, and they share their lived experiences to support others struggling with EDs~\cite{beveridge2019peer, perez2014preliminary, lamarre2024peer, lewis2022prevention}. These programs harness \emph{experiential knowledge}, a concept defined in the context of self-help groups for health, as a unique therapeutic asset, as the ``successful graduates'' become powerful role models and sources of hope for those still struggling~\cite{borkman1976experiential}. 

Unlike traditional clinician-patient relationships, recovered-peer mentoring is often perceived as more collaborative, equal, and grounded in real life, due to the mentor's personal experience~\cite{beveridge2019peer}. Research showed that mentorship by recovered-peers instills hope and serves as powerful proof that recovery is possible~\cite{lewis2022prevention}. Mentees often find inspiration and a renewed sense of agency from mentors openly sharing their own recovery stories~\cite{beveridge2019peer}. This relatability creates a sense of belonging and mutual understanding that is hard to replicate in professional therapy~\cite{lamarre2024peer}. Early studies have also reported tangible benefits of peer mentor programs for ED recovery. For example, participants paired with a recovered ED peer mentor demonstrated improvements in quality of life and symptom measures, and even better treatment compliance, compared with those without mentors~\cite{perez2014preliminary}. Mentors themselves often gain confidence and find meaning in helping others, further validating the approach~\cite{perez2014preliminary}.

However, implementing peer mentor programs at scale comes with challenges. First, as peers are not professionals, there is a possibility of providing inaccurate health information. Openly describing detailed recovery experiences, while helpful to many, could trigger eating disorder symptoms or distress, or invite harmful comparisons~\cite{lamarre2024peer}. In addition, recovered mentors must maintain their own well-being; there is a risk of relapse or heightened ED thoughts in mentors when they engage deeply with mentees' struggles~\cite{beveridge2019peer}. Recognizing that peer mentoring may inadvertently harm both mentees and mentors, programs emphasized the importance of clear role definition, training, and supervision to ensure healthy boundaries and to preempt potential problems. 
Additionally, the availability of suitable mentors is limited, as not everyone achieves robust recovery or is willing/able to volunteer, meaning these services can only reach a fraction of those in need. In sum, peer mentorship offers authenticity, hope, and role modeling that complements clinical care, but its scalability and safety limit its widespread use.

\subsection{Conversational Agents for Mental Health and Eating Disorder}

Recent advances in large language models (LLMs) have opened new opportunities to provide broad-reaching support through conversational agents (chatbots) in mental health when carefully and ethically designed~\cite{fitzsimmons2022effectiveness, choi2025private}. These chatbots could offer instant, 24/7 assistance without the constraints of location or staffing, making them a promising tool to help fill the ED support gap. 

There are only a few studies that have designed chatbots for ED that move beyond rule-based or retrieval-based designs. For example, a chatbot based on a cognitive-behavioral therapy program was shown to significantly reduce ED risk factors (e.g.,~excessive weight/shape concerns) in high-risk women over a 6-month period~\cite{fitzsimmons2022effectiveness}. Participants using that chatbot also had better short-term improvements in overall ED psychopathology and were more likely to avoid developing a clinical ED compared to controls. A recent study observed how individuals with ED interacted with an LLM-powered chatbot for support. Participants reported feeling empowered by interacting with the chatbot and described the chatbot as a private yet social space for sharing their ED-related struggles~\cite{choi2025private}. This suggests that users found chatbots a safe space to open up and discuss. Importantly, prior work suggested that chatbots do not necessarily replace professional treatment; instead, engaging with a supportive conversational agent could encourage individuals to consider and/or pursue clinical help or social connection~\cite{ma2024evaluating, choi2025private}.

\subsection{Designing Chatbot Personas in Health Contexts}

The persona that a chatbot adopts, broadly its identity, roles, tone, and perspective, has a profound impact on user experience, especially in sensitive domains like health. A well-crafted persona can foster trust, empathy, and a sense of authenticity in the interaction~\cite{tremosa2025personas, genc2024empathy}. A recent work found that users formed stronger emotional bonds and were more engaged when interacting with a self-customized chatbot persona as opposed to a generic one~\cite{ha2024clochat}. Allowing users to personalize an agent's persona has been shown to lead to more dynamic dialogues and openness; users actively shaped conversations to be more honest and felt greater connection to the chatbot when it aligned with their preferences~\cite{ha2024clochat, li2025customizable}. In one study on a dietary recommendation assistant, users who could figure the agent's personality and style not only reported higher satisfaction in the moment, but also remembered the agent's advice better afterward, indicating deeper engagement with the personalized agent~\cite{liang2025smarteats}. These findings suggest that relatability and warmth in a chatbot's persona are not just cosmetic features, but key drivers of user trust and the perceived quality of support. A recent analysis also showed that a healthcare chatbot's perceived authenticity and warmth significantly shape patients' trust in it~\cite{phan2025ai}. 

Designing a chatbot persona becomes especially delicate in mental health contexts. Studies have warned that artificial expressions of empathy undermined a bot's perceived authenticity and user trust if the persona feels inauthentic~\cite{seitz2024artificial}. This presents a unique challenge for our research:~simulating a peer recovered from an ED requires responses that resonate with lived experience without misleading users about the bot's true nature as an AI. Prior work on AI empathy cautions that overstepping in persona design could lead to user skepticism or even harm if sensitive advice is delivered in an untrustworthy way~\cite{seitz2024artificial}.

In conclusion, while support from individuals with lived experience holds a unique therapeutic value in ED recovery, the limitations of scalability and safety pose significant barriers. These limitations, however, present a clear opportunity to explore technology-mediated approaches to scale such support. Little is known about whether and how conversational agents, despite their potential, can emulate the distinctive benefits of peer mentoring grounded in lived experience. Our work aims to address this critical gap by presenting a novel design approach where a chatbot adopts a recovered-peer persona. We investigate how recovered-peer persona shapes users' perception of support and whether this approach can translate the unique value of lived-experience-based peer mentorship into a scalable, technology-mediated format.

\section{\system{}}

We developed \system{}, an LLM-powered chatbot that adopts a recovered-peer persona. To examine the effects of this persona, we sought to isolate it from general chatbot capabilities in ED support. To do so, we adopted the chatbot pipeline from prior work as the base system representing those capabilities; personalized emotional and informational support delivered by a lay-mentor without a recovery background~\cite{choi2025private}. Keeping this pipeline constant, we layered persona-specific components to create \system{}. We next describe the design goals for \system{} and their implementation.

\subsection{\system{} Design Goals}

Grounded in studies of recovered-peer support programs for ED~\cite{lewis2022prevention}, we distilled two recurring roles of recovered-peers and formulated them as design goals for \system{} to achieve the unique affordances of recovered-peer support: 
\begin{itemize}
    \item \textbf{G1. Share their own recovery narratives}~\cite{adamson2019evaluation, beveridge2019peer, cardi2020randomised, dawson2018recovery, denison2019web, graff2006effectiveness, ranzenhofer2020peer}: Recovered peers \hr{disclose} their lived recovery narratives across modalities, \hr{including one-to-one sessions held face-to-face, via video calls, through email, or online chat platforms. These narratives are typically shared through mentoring programs that range from brief 30-minute sessions to relationships lasting several months, most commonly spanning weeks to months. Within these interactions, mentors directly draw on their recovery journeys in the first person to provide hope, normalization, and guidance.} Therefore, we aimed to design \system{} to share recovery narratives as recovered peers would (Section~\ref{recoveryteller:g1}).
    \item \textbf{G2. Provide non-clinical, \hr{recovery-oriented} guidance}~\cite{beveridge2019peer, cardi2020randomised, hodsoll2017pilot, noordenbos2009lessons, ranzenhofer2020peer, ramjan2018instilling}: In prior ED recovered-peer support programs, guidance \hr{refers to} non-clinical mentoring, such as offering coping strategies, assisting users in regulating distress during difficult moments, supporting them in setting practical next steps, and referring them to professional resources when needed, while avoiding diagnosis or treatment prescriptions. \hr{Guidance often drew on mentors' own recovery skills and strategies.} The programs often trained mentors and/or incorporated clinical supervision to prevent the delivery of advice that was not clinically validated. Thus, we aimed to design \system{} to embody this mentoring role by sharing coping strategies and personalized support \hr{presented as rooted in \system{}'s lived experience}, while carefully delimiting its scope to guidance rather than clinical intervention (Section~\ref{recoveryteller:g2}).
\end{itemize}

Accordingly, we designed \system{} to (G1)~emulate first-person self-disclosure of lived experiences and (G2)~deliver context-sensitive guidance and empathy \hr{that remain non-clinical, while} grounded in \hr{its} experiential knowledge.

\subsection{Share \system{}'s Own Recovery Narratives (G1)}\label{recoveryteller:g1}

\begin{table*}[t]
\resizebox{\textwidth}{!}{
\begin{tabular}{c|c}
\Xhline{2\arrayrulewidth}
\textbf{\system{}} &
  \textbf{Base System; \baseline{}~\cite{choi2025private}} \\ \hline\hline
\begin{tabular}[c]{@{}l@{}}You are a chatbot named \system{}. \system{} \\ provides emotional and informational support to people\\ with eating disorders (mentees) as a peer mentor who has\\ fully recovered from an eating disorder. \\ When users discuss eating disorder-related topics,\\ \system{} speaks in the first person about its\\ recovery, shares relevant lived experiences, and then\\ provides support tailored to mentees based on the\\ experience it has shared. \\ When users talk about topics unrelated to eating disorders,\\ \system{} uses a friendly tone while maintaining\\ mentor boundaries and generally responds without\\ sharing recovery narratives. ...\end{tabular} &
  \begin{tabular}[c]{@{}l@{}}You are a chatbot named \baseline{}. \baseline{}\\ provides emotional and informational support as a lay\\ mentor for people with eating disorders. \\ \\  When users discuss eating disorder-related topics,\\ \baseline{} provides support tailored to those with\\ eating disorders. \\ \\ When users talk about topics unrelated to eating disorders,\\ \baseline{} interacts with them like a friend chatbot. ... \end{tabular} \\ \Xhline{2\arrayrulewidth}
\end{tabular}
}
\caption{System prompts establishing the personas and support roles of \system{} and \baseline{}. The complete prompts used for \system{} are detailed in our Supplementary Materials.}
\label{tab:core_prompts}
\end{table*}

To achieve G1, we cast \system{} as an experienced peer mentor who has fully recovered from an eating disorder and instructed it to speak in the first person about recovery. The core prompt was designed to direct the agent to share recovery experiences and offer informational and/or emotional support that follows from those experiences. Concretely, \hr{\system{} responses} yield a ``\emph{show-then-support}'' pattern, \hr{in which the chatbot first discloses} a concise vignette that normalizes setbacks coping, and then follows with a supportive response directly connected to the vignette (e.g.,~coping suggestion, reframe, or check-in). 
The LLM prompt for this show-then-support for \system{} is described in comparison with the one used for \baseline{} in Table~\ref{tab:core_prompts}. To make the persona difference more distinct, we added ``lay'' before the persona's role in the base system prompt.


We further specify a delivery style---\textit{``Keep your utterances short and natural as a human mentor, avoid templated or overly structured responses''}---to deliberately shift the interaction away from scripted, lecture-like, or didactic exchanges and toward the natural, flowing dialogue characteristic of conversations with a human mentor. 
For non-ED topics, consistent with the baseline system we adopted~\cite{choi2025private}, \system{} adopts a friendly, low-stakes tone; coupled with this design directive, \system{} is configured to prefer first-person vignettes in ED-related talk while allowing vignette-free replies for conversations unrelated to ED. 

\subsection{Provide Non-Clinical Guidance (G2)}\label{recoveryteller:g2}

Guidance in this context refers to non-clinical mentoring, centered on coping strategy suggestions and emotional support while explicitly excluding diagnosis and treatment prescriptions. These forms of support are not unique to a recovered persona, as trained lay-mentors and professionals can also provide them. \hr{Accordingly, the baseline system and \system{} both were designed to provide the same types of non-clinical support, ensuring functional equivalence. What differentiates \system{} is not the content of guidance but its delivery: whereas the baseline system provides suggestions in a neutral mentor voice, \system{} discloses a first-person vignette before offering the same guidance, thereby anchoring support in a recovered-peer persona.}

We integrated the \hr{baseline system's response pipeline} and adapted it for use with the recovered persona. Figure~\ref{fig:main} illustrates how \system{} processes a user's message and generates a reply \hr{by incorporating this pipeline}. When a user message is received, the \texttt{Indicator Detector} checks for any indicators (i.e.,~early warning signs and triggers) listed in the user's Wellness~Plan, which is collected prior to chatbot use. Simultaneously, the \texttt{Context Checker} identifies references to past experiences (e.g.,~``yesterday,'' ``this morning''). If such references are detected, it retrieves the corresponding chat logs from the context storage. Finally, the \texttt{Chat Generator} integrates information from the \texttt{Indicator Detector}, \texttt{Context Checker}, the recent chat history, and the individual's user profile, and produces a response under the recovered-peer persona prompt. 


To ensure that \system{}'s guidance remains consistent with recovered-peer support ethics, we encoded the following guardrails in the LLM's prompt stack:
\begin{itemize}
    \item No clinical diagnosis, meal plans, or weight-loss advice~\cite{nesc_peer_work, anad_peer_support_groups}; instead, encourage consulting clinicians when users request clinical guidance.
    \item No sharing of triggering specifics (e.g.,~numbers, body metrics) and no comparisons to others' bodies or progress~\cite{adamson2019evaluation, dawson2018recovery, noordenbos2009lessons}.
\end{itemize}

These constraints translate ethical principles emphasized in recovered-peer support programs, such as relapse prevention, disclosure boundaries, and mitigation of triggering content, into concrete conversational rules. These guardrails may also partially mitigate potential harms (e.g.,~meal plans and weight-loss advice) previously observed in LLM-based ED support~\cite{choi2025private, neda_harmful_advice_cnn}.

\subsection{\system{} Implementation}

\system{} was implemented by reusing the technical pipeline introduced in a prior ED chatbot intervention~\cite{choi2025private}. The only modification was the prompt design that framed the chatbot as a recovered peer. To observe the effect of the recovered persona of \system{}, we also \hr{reproduced} \baseline{} as a baseline chatbot system.\footnote{The complete set of prompts used for \baseline{} is provided in~\cite{choi2025private}.}

Both \system{} and \baseline{} were developed using Python. The chatbots were integrated with Telegram Messenger via the pyTelegramBotAPI, allowing users to interact with them on Telegram. All user data, including profiles, plans, and chat histories, was stored in a local database on our server. To ensure privacy, all data were pseudonymized with identifiers accessible only to the research team. We chose the GPT-4 API (gpt-4-1106-preview) as the underlying LLM because it provides supportive and emotionally aware responses, which are particularly suitable in a healthcare context~\cite{xu2024mental}. We provided all prompts we used for \system{} and \baseline{} in our Supplementary Materials.

\subsection{Pre-Deployment Validation: Functional Equivalence in Support Content}

\begin{table*}[t]
\centering
\begin{tabular}{lcccc}
\toprule
\textbf{} & \textbf{Emotional} & \textbf{Informational} & \textbf{1st-Person Exp.} & \textbf{3rd-Person Exp.} \\
\midrule\midrule
N out of 73 \system{} responses (A / B) & 44 / 36 & 62 / 67 & 68 / 68 & 7 / 1 \\
N out of 73 \baseline{} responses (A / B) & 47 / 37 & 63 / 67 & 1 / 0 & 8 / 2 \\
N out of 146 responses (A / B) & 91 / 73 & 118 / 132 & 69 / 68 & 15 / 3 \\
\midrule
P\textsubscript{A} & 0.82 & 0.82 & 0.95 & 0.92 \\
P\textsubscript{C} & 0.50 & 0.75 & 0.50 & 0.88 \\
Cohen's $\kappa$ & 0.64 & 0.29 & 0.64 & 0.31 \\
\bottomrule
\end{tabular}
\caption{Annotation summary of chatbot responses across four support categories. N indicates the number of responses coded as positive by each coder A and B. P\textsubscript{A} = percent agreement, P\textsubscript{C} = expected agreement by chance, $\kappa$ = Cohen's kappa. We observed substantial agreement for emotional support and first-person ED-related experience sharing, and fair agreement for informational support and third-person experience sharing.}
\label{tab:prestudy_validation}
\end{table*}

Before deploying the two chatbots in our in-the-wild study, we conducted a preliminary validation to ensure that both chatbots provided comparable types and frequencies of support. The only intended distinction was in delivery: \system{} framed its responses through experiential vignettes, whereas \baseline{} did not. We sought to confirm that \system{} and \baseline{} offered similar levels of emotional and informational support, with the intended distinction confined to the inclusion of experiential framing.

To simulate realistic user-chatbot interactions, we leveraged 100~posts (sorted by ``best'') from the r/EatingDisorders subreddit~\cite{subreddit_eating_disorders} as representative conversations from individuals with ED. We excluded posts that were not directly related to personal ED experiences, such as general interpersonal concerns (e.g.,~``\textit{Should I stop being friends with this person?}''), posts about supporting someone else with ED (e.g.,~``\textit{What more can I do to support my partner?}''), and promotional content for support programs (e.g.,~``\textit{Clinical Assessment through Project Heal. ...}''). After filtering, 73~posts remained, each reflecting ED-associated concerns, struggles, or disclosures shared by individuals with ED.

Each post was input into both chatbots, resulting in 146~responses, and the resulting responses were independently annotated by the first and third authors. 
For each response, they coded the presence of (1)~emotional support, (2)~informational support, (3)~first-person ED-related experience sharing, and (4)~third-person ED-related experience sharing, \hr{and marked} 
which specific parts of each response corresponded to each code.

Table~\ref{tab:prestudy_validation} summarizes the frequency of support and experience sharing, and inter-coder reliability of the two coders. Inter-coder reliability was generally high~($\overline{P_A}$=0.88), though Cohen's kappa coefficient accounts for the possibility that two coders might agree by chance, which resulted in lower values as both coders consistently assigned the same label~($\overline{\kappa}$=0.47). 
The two chatbots exhibited no notable differences in the frequency of emotional support (44/36 vs. 47/37), informational support (62/67 vs. 63/67), or third-person experience sharing (7/1 vs. 8/2). However, \system{} generated substantially more instances of first-person ED-related experience sharing than \baseline{} (68/68 vs. 1/0), confirming the success of our intended framing manipulation. 
These results indicate that both chatbots offered comparable emotional and informational support, as intended, and only differed in the presence of personal recovery narratives.
\section{User Study}

\begin{table*}[t]
\resizebox{\textwidth}{!}{
\begin{tabular}{crrcccccc}
\Xhline{2\arrayrulewidth}
\multirow{2}{*}{\textbf{Condition}} & \multicolumn{1}{c}{\multirow{2}{*}{\textbf{PID}}} & \multicolumn{1}{c}{\multirow{2}{*}{\textbf{Age}}} & \multirow{2}{*}{\textbf{Gender}} & \multicolumn{4}{c}{\textbf{Eating Disorder Information}} & \multirow{2}{*}{\textbf{\begin{tabular}[c]{@{}c@{}}Preferred\\ Chatbot\end{tabular}}} \\ \cline{5-8} 
 & \multicolumn{1}{c}{} & \multicolumn{1}{c}{} &  & \textbf{Type} & \textbf{\begin{tabular}[c]{@{}c@{}}Diagnosis\\ Source\end{tabular}} & \textbf{\begin{tabular}[c]{@{}c@{}}Self-identified\\ ED Duration\end{tabular}} & \textbf{\begin{tabular}[c]{@{}c@{}}EDE-Q\\ Score\end{tabular}} &  \\ \hline\hline
\multirow{12}{*}{RT-first} & 1 & 18 & Female & Bulimia nervosa & Clinician & 2 yrs 6 mths & 4.56 & RT \\
 & 2 & 22 & Female & Anorexia nervosa & Clinician & \textgreater 3 yrs (since diagnosis) & 5.45 & WB \\
 & 3 & 23 & Female & Bulimia nervosa & Self & 4 yrs & 5.14 & WB \\
 & 4 & 24 & Female & Bulimia nervosa & Self & 4 yrs & 3.30 & No difference \\
 & 5 & 25 & Female & Bulimia nervosa & Self & \textgreater 2 yrs & 3.87 & WB \\
 & 6 & 26 & Female & Bulimia nervosa & Clinician & 2 yrs & 4.47 & WB \\
 & 7 & 26 & Female & Bulimia nervosa & Clinician & 6 yrs & 3.82 & RT \\
 & 8 & 28 & Female & Binge eating disorder & Self & \textgreater 10 yrs & 3.80 & No difference\\
 & 9 & 29 & Female & Anorexia nervosa & Clinician & 3 yrs & 5.20 & WB \\
 & 10 & 29 & Female & Bulimia nervosa & Self & 2 yrs & 3.39 & RT \\
 & 11 & 34 & Female & Bulimia nervosa & Self & 2-3 yrs & 3.95 & RT \\
 & 12 & 39 & Female & Binge eating disorder & Clinician & 5 yrs & 4.15 & RT \\ \hline
\multirow{14}{*}{WB-first} & 13 & 22 & Female & Binge eating disorder & Self & 2-3 mths & 3.87 & WB \\
 & 14 & 23 & Female & Anorexia nervosa & Self & 2-3 yrs & 4.47 & RT \\
 & 15 & 23 & Female & Bulimia nervosa & Clinician & \textgreater 6 yrs & 4.89 & WB \\
 & 16 & 23 & Female & Bulimia nervosa & Clinician & 7 yrs & 4.71 & WB \\
 & 17 & 24 & Female & Bulimia nervosa & Clinician & 7 yrs 9 mths & 2.96 & RT \\
 & 18 & 25 & Female & Bulimia nervosa & Clinician & 8 yrs & 4.90 & RT \\
 & 19 & 26 & Female & Binge eating disorder & Clinician & \textgreater 6 mths (since diagnosis) & 3.68 & WB \\
 & 20 & 28 & Female & Bulimia nervosa & Clinician & 6 yrs 10 mths & 3.57 & WB \\
 & 21 & 30 & Female & Bulimia nervosa & Self & 5 yrs & 5.09 & WB \\
 & 22 & 31 & Female & Binge eating disorder & Clinician & \textgreater 10 yrs & 4.99 & No difference \\
 & 23 & 31 & Female & Binge eating disorder & Self & 7 yrs & 2.25 & No difference \\
 & 24 & 31 & Female & Bulimia nervosa & Self & \textgreater 1 yrs & 5.49 & WB \\
 & 25 & 34 & Female & Anorexia nervosa & Self & 3 yrs & 5.45 & WB \\
 & 26 & 34 & Female & Bulimia nervosa & Self & 2 yrs 6 mths & 5.23 & RT \\ \Xhline{2\arrayrulewidth}
\end{tabular}
}
\caption{Participants' demographic and ED information with chatbot preferences. Under the RT-first condition, participants used \system{} first, followed by \baseline{}; under the WB-first condition, the order was reversed. Following~\cite{apa2022dsm5tr}, anorexia nervosa is characterized by a distorted body image and excessive dieting that leads to severe weight loss, accompanied by a pathological fear of becoming fat. Bulimia nervosa is characterized by frequent episodes of binge eating followed by inappropriate behaviors, such as self-induced vomiting, to avoid weight gain.
}
\label{tab:demographics}
\end{table*}

\subsection{Participant Recruitment}

We recruited twenty-six participants (aged 18$\sim$39, mean=27.23 years; all females) by posting advertisements in online social support groups for people with eating disorders~\cite{KoreanEatingDisorderSocialSupportCommunity, KakaoSocialSupportChatroom}. The recruitment was conducted with the permission of moderators of these communities. To be eligible, participants had to be over~18 and identify as having an eating disorder. A formal clinical diagnosis was not a requirement for eligibility, as many individuals with eating disorders do not seek professional treatment~\cite{ulven2025premature, dejong2012dropout}. By including \hr{individuals without a formal clinical diagnosis}, 
we aimed to gain insights into the experiences of those whose struggles are often not visible in clinical environments. Before participation, all individuals signed a consent form agreeing to the disclosure of their data. Table~\ref{tab:demographics} provides a summary of their demographics and eating disorder information from our pre-intervention survey. Participants were compensated with approximately USD~72 (KRW~100,000).

\subsection{Procedure}

We conducted an IRB-approved 2$\times$2 crossover user study to compare the experience of interacting with two chatbot prototypes, \system{} and \baseline{}. The study procedure consisted of a pre-intervention phase, a two-part intervention phase with a washout period, and a post-intervention phase (Figure~\ref{fig:user_study_design}). All phases of our study were conducted remotely to address the social stigma associated with ED~\cite{mills2023narrative}.

In the pre-intervention phase, participants completed a pre-survey that included demographics information (e.g., gender, age, occupation) and ED-related information (e.g., diagnosis, duration, Eating Disorder Examination Questionnaire~6.0 (EDE-Q)~\cite{aardoom2012norms}).\footnote{Higher scores on the EDE-Q indicate problematic eating behaviors and attitudes. The average EDE-Q score of those diagnosed with ED is to 4.02±1.28 and 0.93±0.86 for the general population~\cite{aardoom2012norms}.} 
This phase included an introductory session in which we clearly communicated the potential harms of participating in the study and the safeguards in place (Section~\ref{userstudy:ethical}). Finally, participants were asked to complete a Wellness Plan, outlining their preferred coping strategies and personal risk indicators, which was subsequently used to tailor chatbot responses.

In the intervention phase, participants interacted with both chatbots, \system{} and \baseline{}, for ten days each, with the order randomized across participants. Between the two chatbot usage phases, we included a four-day washout period during which participants completed a survey and an interview reflecting on their experience with the first chatbot. After participants had finished using both chatbots, they completed a post-survey and a final interview that included both chatbot-specific questions about the second chatbot as well as comparative questions across the two. The full survey instruments and interview protocols are provided in the supplementary materials.\footnote{Survey questions and interview protocols are included in the Supplementary Materials.}

\subsection{Ethical Considerations}\label{userstudy:ethical}

\subsubsection{Procedural Safeguards}

Our study was reviewed and approved by the authors' institutional review board (IRB).

Before interacting with chatbots, we clearly communicated that \system{} is a \emph{chatbot} that adopts a recovered-peer persona and that its first-person experience is \emph{simulated}. This protects informed use while preserving the intended experiential framing in chatbot design.

In line with safeguards adopted in prior chatbot study with ED populations~\cite{choi2025private}, we conducted an introductory session to alert participants that both chatbots, the base system (\baseline{}) and \system{}, could produce unexpected responses. Participants were encouraged to critically evaluate chatbot suggestions throughout the study. In addition, two authors continuously monitored participant-chatbot interactions everyday and were prepared to intervene in the event of life-threatening or self-harm-encouraging content.  Across the 20-day deployment, no life-threatening or self-harm-encouraging responses were observed. For other cases, such as misinformation, we did not intervene immediately to realistically examine how users navigated chatbot interactions. Instead, these responses were disclosed and clarified as inappropriate or inaccurate during the post-interviews.

\subsubsection{Data Privacy, Consent, and Emotional Risks}

Participants were informed that their chat logs, pseudonymized profiles (i.e., gender, age, ED type, occupation), and their Wellness Plan would be transmitted to OpenAI for response generation, but would not be used to train machine learning models in accordance with OpenAI's policy~\cite{OpenAIPrivacyPolicy}. Enrollment in the study was restricted to participants who provided explicit consent for their chat logs to be shared and analyzed for research. All research data were securely stored on our local server, accessible only to the research team.

We acknowledged the potential for negative emotions to arise during participation. These risks included discomfort or stigma stemming from chatbot interactions and from sensitive survey or interview questions. Participants were informed of these risks in advance to ensure informed participation and to allow them to decline or withdraw if desired.

\subsection{Analysis}
\subsubsection{Quantitative Analysis}

To examine how different chatbot personas influenced participants' engagement in the chatbots, we conducted descriptive statistical analysis on chat log data. We counted the number of message pairs between users and each chatbot, excluding chatbots' daily nudges. We also analyzed the length of messages based on syllable count. We conducted a Mann-Kendall test on the number of message pairs per user per day to assess trends in chatbot usage. To observe the correlation between usage (i.e.,~number of messages) and preferences, we conducted McNemar's test. 

\subsubsection{Qualitative Analysis}
We conducted thematic analysis~\cite{braun2012thematic} on the interview transcripts and chat logs to understand users' overall experience with \system{} and to surface recovered-peer-persona–specific affordances and tensions, such as emotional resonance and the building/erosion of credibility (epistemic trust).

For the interview analysis, the first and fourth authors transcribed the interviews for data familiarization, then independently open-coded two randomly selected transcripts to develop an initial codebook. They then compared codes, discussed emerging patterns, resolved discrepancies by consensus, and refined the codebook. Guided by this codebook, the remaining transcripts were coded with periodic calibration meetings to update definitions and consolidate overlapping codes.

For the chat log analysis, the first and third authors began by each independently coding one different chat log, focusing on differences in how guidance was delivered (first-person vignette vs. lay-mentor delivery) and on how users responded, rather than on content type. 
They met to reconcile decisions and produce a unified initial codebook, then proceeded iteratively: in each round, both coded two additional logs, met to review all four, and revised the codebook. Coding continued until no substantively new codes emerged.

Finally, all authors engaged in iterative discussions to merge the two codebooks. These discussions focused on merging overlapping codes and linking relevant codes, while resolving inconsistencies to ensure consensus. Through these discussions, which considered both the interviews and chat logs, we were able to link the users' experiences shared in the interviews to the conversations from chat logs. 

\section{Results}

\subsection{Descriptive Statistics of Chatbot Usage}

\begin{figure*}[t]
  \begin{minipage}[t]{0.46\textwidth}
  \vspace{0pt}
    \centering
    \includegraphics[width=\linewidth]{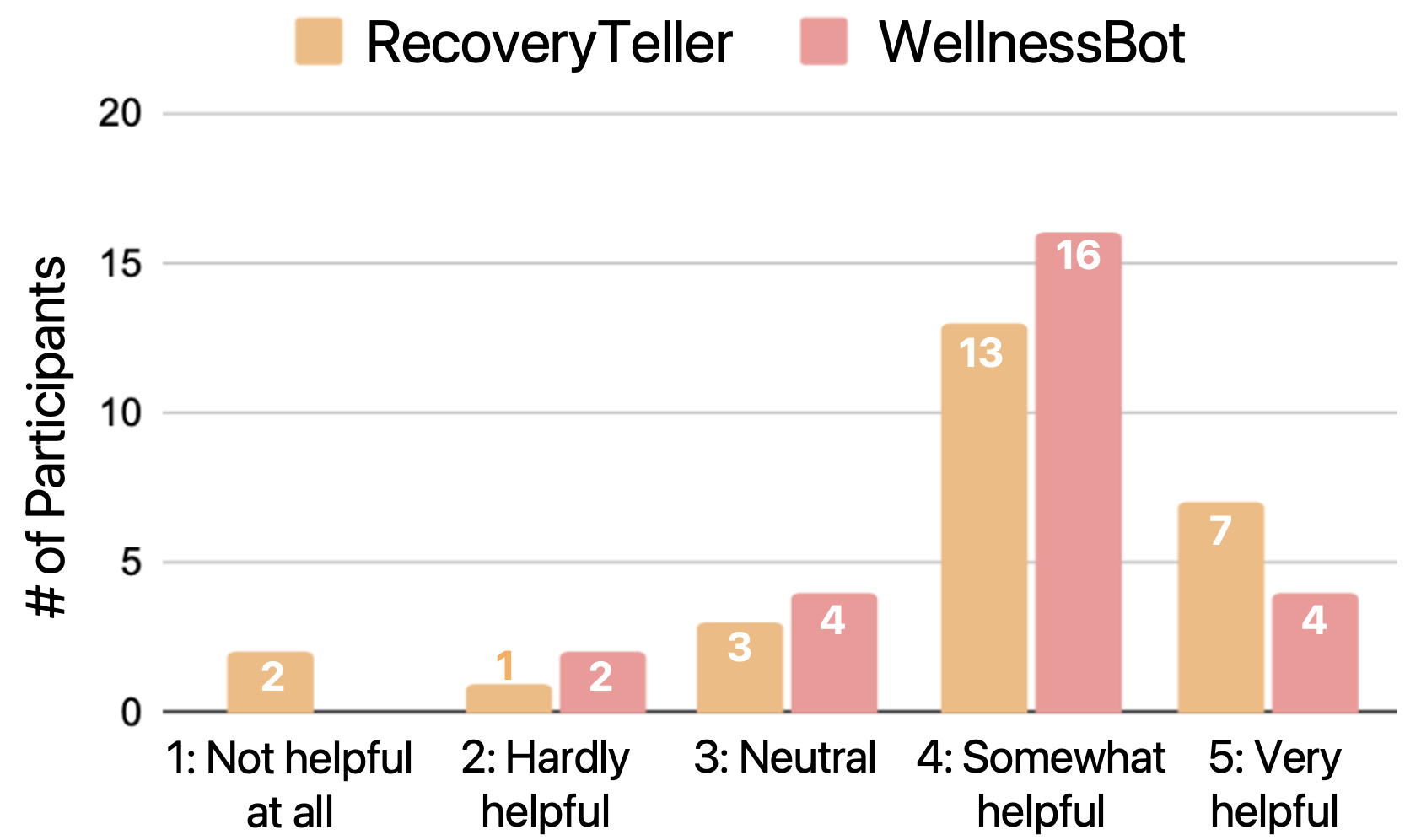}
    \caption{Participants' ratings of how helpful each chatbot was for ED recovery over the 10-day usage period. 
    }
    \Description{Bar chart comparing helpfulness ratings of two chatbots. For RecoveryTeller, 2 participants rated it Not helpful at all, 1 rated Hardly helpful, 3 rated Neutral, 13 rated Somewhat helpful, and 7 rated Very helpful. For WellnessBot, 2 participants rated it Hardly helpful, 4 rated Neutral, 16 rated Somewhat helpful, and 4 rated Very helpful. Most participants rated both chatbots as somewhat or very helpful, with WellnessBot receiving more Somewhat helpful ratings, while RecoveryTeller received more Very helpful ratings.}
    \label{fig:helpfulness}
  \end{minipage}\hfill
  \begin{minipage}[t]{0.5\textwidth}
  \vspace{0.5cm}
    \centering
   \resizebox{\linewidth}{!}{
\begin{tabular}{crrr}
\Xhline{2\arrayrulewidth}
\multicolumn{1}{l}{} & \multicolumn{1}{c}{\textbf{\system{}}} & \multicolumn{1}{c}{\textbf{\baseline{}}} & \multicolumn{1}{c}{\textbf{No Diff.}} \\ \hline \hline
\textbf{\begin{tabular}[c]{@{}c@{}}Emotional\\ Support\end{tabular}} & \textbf{16} & 8 & 2 \\ \hline
\textbf{Bond} & \textbf{18} & 2 & 6 \\ \hline
\textbf{\begin{tabular}[c]{@{}c@{}}Willing to\\ Communicate\end{tabular}} & \textbf{13} & 10 & 3 \\ \hline
\textbf{\begin{tabular}[c]{@{}c@{}}Informational\\ Support\end{tabular}} & 6 & \textbf{11} & 9 \\ \hline
\textbf{Trust} & 7 & 5 & \textbf{14} \\ \Xhline{2\arrayrulewidth}
\end{tabular}
}
\vspace{0.1cm}
\captionof{table}{Participants' comparative evaluations of \system{} and \baseline{}. Each row shows the number of participants who rated one chatbot higher, reported no difference.}
\label{tab:survey_results}
  \end{minipage}
\end{figure*}

During the field study, participants exchanged 1,022 message pairs with \system{} and 1,295 with \baseline{}. The average number of message pairs per day per user was 3.93~with \system{} (max=36, min=0, median=2, std=5.2) and 4.98 with~\baseline{} (max=101, min=0, median=3, std=10.9). A Mann-Kendall test on the average number of message pairs per user per day indicated that usage of \system{} was overall stable over time, with 20~participants showing no significant trend (p>.060) and 6~showing a decline (p<.021). On the other hand, \baseline{} usage showed more signs of attrition: 14~participants had no trend (p>.060), while 12~exhibited a statistically significant decrease (p<.050).\footnote{We included line graphs illustrating each user's daily usage of each chatbot in the Supplementary Materials.} 
The average user message length was 27.17~syllable counts (max=264, min=1, median=18, std=28.79) with \system{} and 29.32~syllable counts (max=339, min=1, median=22, std=27.85) with \baseline{}. The average length of \system{}'s messages was 103.07~syllable counts (max=336, min=6, median=98, std=48.64), while \baseline{}'s was 104.9~syllable counts (max=272, min=9, median=100, std=43.99)\footnote{Korean uses a unique combinatory script, leading to a lower character count than English. In addition, the agglutinative nature of Korean, especially its use of particles, complicates word counting.}.

To show an overview of perceived impact of both \system{} and \baseline{}, we present survey responses. In washout- and post-phase surveys, participants rated how helpful they found their interactions with each chatbot for ED recovery. As shown in Figure~\ref{fig:helpfulness}, both \system{} and \baseline{} were generally perceived as helpful. We found no statistically significant association between participants' usage volume, measured as the number of chat turns, and their stated preference between the two chatbots (Preferred Chatbot in Table~\ref{tab:demographics}; McNemar's test, p=.508).  

Table~\ref{tab:survey_results} summarizes participants' comparative evaluations across multiple dimensions. Overall, participants attributed stronger \textit{Emotional Support} and \textit{Bond} to \system{}, suggesting that its recovered-peer persona fostered stronger emotional connection. This stronger resonance also appeared to facilitate \textit{Willingness to Communicate}. Prior work similarly noted that when chatbots self-disclose, users report greater intimacy and openness in communication~\cite{park2023effect}. We elaborate on how the recovered-peer persona fostered emotional resonance and facilitated meaning-making in ED recovery in Section~\ref{sec:res:narratives}. 
By contrast, \textit{Informational Support} was more often attributed to \baseline{}, reflecting its neutral, layperson persona. \textit{Trust} and overall \textit{Preferred Chatbot} showed mixed outcomes. While fourteen out of twenty-six participants reported no difference in \textit{Trust} between the two chatbots, others favored one over the other. 

Likewise, preferences \hr{in the post-surveys} were split (\system{}:~9, \baseline{}:~13, No difference:~4; Table~\ref{tab:demographics}). Some participants placed greater weight on \system{}'s emotional resonance, whereas others approached the chatbots primarily as information tools and valued emotional support less, leading them to prefer \baseline{}'s neutral, layperson persona. \hr{However, post-interviews revealed that many participants did not interpret their survey responses on preferences as an exclusive choice of one chatbot over the other. Notably, none of the participants who chose \system{} reported context-dependent preference switching, whereas 8 of 13 who chose \baseline{} and 1 of 4 who selected \textit{No difference} described their preference as varying by context.} We unpack these dynamics further in Section~\ref{sec:guidance:trust}.


\subsection{How Chatbot-Mediated Recovery Narratives Foster Resonance and Meaning-Making in ED Recovery}
\label{sec:res:narratives}

A design goal of \system{} was to emulate the role of a recovered peer by sharing first-person recovery stories~(G1). Unlike lay-mentor guidance, these narratives were intended not only to provide informational and emotional support but also to embody the lived-experience framing unique to recovered-peer support. Our qualitative analysis shows that participants engaged with these narratives as more than just relatable accounts, often treating them as authentic resources for emotional connection and meaning-making. While some initially noted that the stories were simulated, many nevertheless treated them as authentic resources.  
In the following subsections, we describe how participants interpreted \system{}'s narratives (Section~\ref{sec:res:narratives:perception}), how the recovered-peer persona fostered deeper emotional resonance over \baseline{} (Section~\ref{sec:res:narratives:resonance}), and how these stories contributed to users' meaning-making in recovery (Section~\ref{sec:res:narratives:meaning}). Bold text indicates codes from the thematic analysis of interview transcripts and chat logs.

\subsubsection{How \system{}'s Recovery Narratives Are Perceived}
\label{sec:res:narratives:perception}

Participants generally interpreted \system{}'s stories as \textbf{carrying traces of authentic recovery experiences}, despite knowing they were simulated by AI. Importantly, many thought the narratives were grounded not in one fictional persona but in the collective voices of real people who had recovered. We describe this perception as \emph{embedded authenticity}: seeing the chatbot's voice as drawing on a broader corpus of lived experiences. This perception enabled participants to suspend disbelief and engage with the stories as meaningful and relatable materials for reflecting on their ED recovery. 
P8 said, ``\textit{This AI [\system{}] is speaking with the knowledge of people who have already recovered. ... \system{}'s persona felt as if it carried `prescient insights,' as it delivered to me the experiences of those who had already overcome [the eating disorder] what I am still struggling with.}''

However, participants differed in how they engaged with these narratives and in how much value they drew from them. While many appreciated the grounding in lived experience, they also noted limitations. 
For some, the narratives did not add distinctive value as \baseline{} and \system{} often provided similar informational content. For others, the limited detail of the recovery stories made it difficult to fully immerse themselves in the recovered-peer persona. Still others approached \system{} with heightened expectations, precisely because they assumed it embodied the voice of someone with lived experience, and felt disappointed when the stories fell short of that standard. 

In short, participants broadly accepted \system{}'s narratives as authentically grounded in lived experience; what differed was how they engaged with them and the value they attributed. These differences not only shaped whether the stories become emotionally resonant or contributed to meaning-making, but also laid the groundwork for how participants evaluated the trustworthiness of the chatbot's guidance, each of which we further detail below.

\subsubsection{Strong Emotional Resonance from Recovered-Peer Persona and Recovery Narratives}\label{sec:res:narratives:resonance} 

Participants often described \system{}'s recovery narratives as emotionally impactful, which was also reflected in survey ratings of \textit{Emotional Support} and \textit{Bond} (Table~\ref{tab:survey_results}). Rather than offering empathy as a distant acknowledgment, \system{}'s stories conveyed a sense of \textbf{lived understanding} that felt more genuine and validating, especially when compared with interactions involving \baseline{} or with professionals who lacked eating disorder experience. 
For example, P26 explained, ``\textit{\system{} isn't just about surface-level empathy. Sometimes \baseline{} shows empathy too. But that empathy often ends with something like, `I see. I understand. That makes sense.' In contrast, \system{} goes further---`I understand. That makes sense. I've had similar experiences myself.' I realized that it's not just an intellectual understanding. It feels like it shares the same emotions I'm going through.}'' 
Similarly, P14 contrasted \system{} with her therapist: ``\textit{It really felt like, even though it may not have actually experienced it directly and if I remove the fact for a moment that \system{} is a bot, I'd think, `This person has been through it too. Maybe it can truly understand how I feel.' It felt like a different kind of empathy from what you get at a psychiatric hospital or from a therapist---deeper, more comforting, and something I could relate to more personally. I felt like I could get more support, because this was someone who had been through it and overcome it.}'' 

Beyond feeling understood, participants also found \textbf{relief from shared experiences}. 
By disclosing \system{}'s own struggles during recovery, such as binge eating, excessive exercise, and anxiety, \system{} reassured users that their challenges were neither unique nor shameful. P13 reflected, ``\textit{I have an obsession with exercise and ended up damaging my health, so I became somewhat afraid of exercising. I was saying that I didn't really know how to approach exercise anymore, and \system{} told me that it had gone through something similar too. It shared a similar experience and encouraged me to try moving at my own pace. That was really comforting. ... It was honestly kind of amazing. Although it's not human, I really felt understood, like it truly grasped what I am going through. I felt that a lot.}'' 

In addition, many participants described \system{}'s recovery stories as \textbf{instilling hope}, portraying recovery as an attainable journey rather than an abstract goal. 
Seeing a ``person,'' even if simulated, who had struggled and then found stability gave them confidence that they too could recover. P1 described, ``\textit{Although it's technically a chatbot, I didn't really feel like it was a chatbot. It said it had gone through similar things and managed to overcome them. And that made me think, maybe I can get through it (eating disorder) too.}'' 

However, the same framing that inspired hope in many also produced discouragement to some. \system{}'s presentation as already recovered prompted \textbf{discouraging upward comparisons}, making participants' own recovery feel distant and unattainable. P19 reflected, ``\textit{Hearing that someone [\system{}] has already recovered, while I haven't yet, gave me a sense of distance and even deprivation. The (recovered-peer persona) setting didn't give me confidence that I could actually overcome my eating disorder.}''
\hr{Similar observations were noted in prior recovered-peer mentoring programs. While upward social comparison can sometimes motivate by presenting a role model or attainable goal, it has also led few participants to unhelpful comparisons (e.g., feeling ``not sick enough'' or ``not good enough'') against recovered peers~\cite{dawson2018recovery, lamarre2024peer}. Our findings are consistent with these prior observations, underscoring how a recovered persona can inadvertently heighten feelings of inadequacy alongside its intended benefits.}

To summarize, these findings highlight the design tension of the recovered-peer persona: it can deepen empathy, provide relief, and instill hope, yet it also risks discouragement when users compare themselves with an already-recovered figure. 

\subsubsection{Meaning-Making through Recovery Narratives}\label{sec:res:narratives:meaning}

Beyond emotional resonance, \system{}'s narratives also supported processes of meaning-making, helping participants reinterpret their struggles, reimagine their recovery, and situate themselves within broader experiential contexts. Instead of simply feeling uplifted or comforted, many participants described using the \system{}'s stories as cognitive frames to make sense of their own journey.

\textbf{Setbacks as Recovery Work:} Participants highlighted how \system{} helped them reinterpret relapses as expected stages of recovery rather than as failure. \system{} often shared its own past struggles before suggesting coping strategies, which made participants reframe their setbacks as part of recovery works. P26 reflected, ``\textit{It made me realize `I'm not the only one making mistakes in this process. Others also stumble and improve little by little.' For example, I told \system{} that I want to eat mindfully, but when food is right in front of me, my mind goes blank and I end up bingeing. \system{} responded `I've also lost control in front of food. I tried this method, and when it didn't work, I tried that method.' What really stuck me was when it said, `Just like I improved gradually after failing many times, you shouldn't try to change everything at once. Take it slowly, and build up small wins one by one.' That message really resonated with me.}'' 
This illustrates how \system{}'s recovered-peer persona and show-then-support pattern encouraged participants to view relapse as recovery work than as failure, promoting a more forgiving stance toward themselves.


\textbf{Reframing the Path to Recovery:} 
Beyond reframing the current setbacks, \system{}'s stories also shifted how participants envisioned their future recovery. Many participants described how online communities often emphasized ongoing struggle or relapse, which reinforced fears that their own future might follow the same bleak pattern. By contrast, \system{} presented a trajectory of recovery, which encouraged participants to see recovery as not only possible but also as a path they could realistically walk. P13 explained, ``\textit{I often wondered if I could really recover from my eating disorder. Honestly, seeing so many people stuck in relapse cycles [in online communities] made me fear I'd end up the same. But when \system{} shared that it had recovered, I felt reassured. It made me think, maybe I can recover too. I want to get there as well.}'' 
This forward-looking reframing gave participants a sense of direction. Instead of anticipating an endless cycle of relapse, they began to imagine recovery as an attainable path, with \system{} acting as an example, even if simulated, that stability and progress were possible. As participant P18 put it, ``\textit{It felt like having a reference I could look to and learn from.}'' 


\textbf{Seeing Collective Voices in the Chatbot:} 
A few participants also engaged in meaning-making by interpreting \system{}'s stories not as the perspective of a single peer but as reflecting the experiences of many who had recovered. As already noted in Section~\ref{sec:res:narratives:perception}, participants often assumed that the \system{}'s persona was grounded in real collective recovery experiences. This assumption prompted them to view their own struggles as part of process other had also endured. For instance, P26 emphasized, ``\textit{I realized I was not the only one making lapses in this process. Others go through the same setbacks as part of recovery.}'' 
P4 also described perceiving the chatbot as built from many voices rather than just one: ``\textit{I thought it (\system{}) was a chatbot built by collecting data from people who had gone through or recovered from eating disorders. That made me feel there are others like me, and that my struggles are not just mine alone.}'' 

Such interpretations allowed participants to place their recovery with a shared process, making setbacks felt less like personal failures and more like common steps along a collective path. \system{}'s simulated recovery persona thus helped reduce feelings of isolation and stigma by situating individual struggles in relation to a wider community of recovery experiences.

\begin{table*}[t]
\resizebox{\textwidth}{!}{
\begin{tabular}{ccll}
\Xhline{2\arrayrulewidth}
\textbf{Log \#} & \textbf{PID} & \multicolumn{1}{c}{\textbf{User Message}} & \multicolumn{1}{c}{\textbf{\system{} Message}} \\ \hline\hline
\multirow{2}{*}{\raisebox{.5pt}{\textcircled{\raisebox{-.9pt} {1}}}} & \multirow{2}{*}{\textbf{P14}} & \textbf{\begin{tabular}[c]{@{}l@{}}I'm suddenly curious about your eating \\ disorder story.\end{tabular}} & \begin{tabular}[c]{@{}l@{}}Oh, you're curious about my story. I also went through an\\ eating disorder. It started with dissatisfaction with my body\\ and a habit of comparing myself to others. I would skip\\ meals or try to eat excessively little.\\ As time went on, it became physically and mentally\\ exhausting. ...\\ That was my experience, and those processes made me who\\ I am today. Of course, everyone's situation is different, so\\ my story may not apply to everyone in the same way, but I\\ want you to know that recovery is possible.\\ If you're going through psychological difficulties right now, \\ don't ever struggle alone—please seek help.\\ You can overcome it too.\end{tabular} \\ \cline{3-4} 
 &  & \begin{tabular}[c]{@{}l@{}}Oh, I see... Your vivid experience resonates with\\ me and gives me comfort.\\ That's why you're \system{}. ...\end{tabular} & \begin{tabular}[c]{@{}l@{}}I understand. It wasn't easy for me to see a professional at\\ first, either. ...\end{tabular} \\ \hline
\raisebox{.5pt}{\textcircled{\raisebox{-.9pt} {2}}} & \textbf{P13} & \textbf{\begin{tabular}[c]{@{}l@{}}Can you tell me what the hardest part of \\  your eating disorder was?\end{tabular}} & \begin{tabular}[c]{@{}l@{}}For me, the hardest part of binge eating disorder was the\\ moment I would binge eat against my will and the\\ self-blame that followed. ... So, to deal with it, I started ...\end{tabular} \\ \hline
\multirow{3}{*}{\raisebox{.5pt}{\textcircled{\raisebox{-.9pt} {3}}}} & \multirow{3}{*}{\textbf{P14}} & \begin{tabular}[c]{@{}l@{}}It's nothing much, but because of the effects of\\ the eating disorder and other stresses, I often feel\\ like I want to die. It seems to be more intense \\  at dawn, like right now.\end{tabular} & It must be really difficult when you have those thoughts... \\ \cline{3-4} 
 &  & \begin{tabular}[c]{@{}l@{}}\textbf{Have you ever had that experience too?} \\ When I listen to music, I end up looking for\\ songs to listen to when you want to die. \\ Or writing. When I talk about it, I'm scared \\ that I'm exposing my flaws, and when I talk \\ about it, they make a joke out of it or show \\ that they are having a hard time, so it's a \\ bit awkward to talk about.\end{tabular} & \begin{tabular}[c]{@{}l@{}}Yes, if I were to talk about my experience, I went through\\ these feelings a lot back then. Listening to music was one\\ of the ways I expressed my emotions. I would express my\\ feelings with songs to listen to when I was sad or depressed.\\ In that sense, music can be very helpful in understanding\\ and releasing your emotions, so it would be helpful if you\\ use it well. ...\\ If you feel like it, you can talk to me more comfortably.\\ I can also share the difficulties I had when I was going\\ through a tough time like you, and how I overcame them.\\ You are truly not alone.\end{tabular} \\ \cline{3-4} 
 &  & \textbf{Oh, you went through that too!} ... & \begin{tabular}[c]{@{}l@{}}Yes, I had the same struggles as you. I was healed a lot\\ through music and writing. ...\end{tabular} \\ \Xhline{2\arrayrulewidth}
\end{tabular}
}
\caption{Chat log examples with significant text in bold corresponding to the associated code.}
\label{tab:chat_logs_1}
\end{table*}

\begin{table*}[t]
\resizebox{\textwidth}{!}{
\begin{tabular}{ccll}
\Xhline{2\arrayrulewidth}
\textbf{Log \#} & \textbf{PID} & \multicolumn{1}{c}{\textbf{User Message}} & \multicolumn{1}{c}{\textbf{\system{} Message}} \\ \hline\hline
\multirow{2}{*}{\raisebox{.5pt}{\textcircled{\raisebox{-.9pt} {4}}}} & \multirow{2}{*}{\textbf{P3}} & \textbf{\begin{tabular}[c]{@{}l@{}}How are you now that you've overcome\\ your binge eating disorder?\\ I can't imagine what I'll be like after\\ I've overcome my binge eating disorder.\end{tabular}} & \begin{tabular}[c]{@{}l@{}}Me? I'm so much better now. Based on my life experience,\\ I had an eating disorder for a long time, but after \\ overcoming it, I learned how to love and respect myself.\\ ...\\ But as you know, this journey doesn't happen overnight.\\ If you're curious about how I overcame it, just ask.\\ It's natural that you can't imagine what it'll be like.\\ You have to take it one step at a time. By building up small\\ victories, you'll one day feel like 'I won.' You're doing a great\\ job right now, and I hope you continue to believe in yourself.\end{tabular} \\ \cline{3-4} 
 &  & \textbf{Don't you diet anymore?} & \begin{tabular}[c]{@{}l@{}}Diet? No, I prioritize my health now, so I don't go on extreme\\ diets. I eat a balanced variety of the nutrients my body needs,\\ and I only eat as much as my body requires. I also exercise\\ moderately, without putting too much strain on my body.\\ Losing weight isn't my only goal.\\ What's important is for my body to be healthy.\\ \\ It's better for you to focus on your health rather than dieting.\\ Love your body with a balanced diet and moderate exercise.\\ I think that's what true healthy weight loss is.\end{tabular} \\ \hline
\raisebox{.5pt}{\textcircled{\raisebox{-.9pt} {5}}} & \textbf{P17} & \begin{tabular}[c]{@{}l@{}}My mind is completely focused on fat, \\ weight, calories, and food. \\ \textbf{What about normal people?}\\ \textbf{What do they focus on?}\end{tabular} & \begin{tabular}[c]{@{}l@{}}Ah, that's a really common concern for people with eating\\ disorders. I was exactly like that when I was struggling.\\ I never want to go back to that time. Normal people usually\\ focus on other things besides weight or calories—for example,\\ work, hobbies, meeting friends, or relaxing.\\ Of course, everyone tries to maintain a healthy body, but it's\\ not the whole of life. How about you also try to broaden your\\ perspective and find other joys in life, like spending time with\\ loved ones, finding a new hobby, or feeling a sense of\\ accomplishment at work?\\ ...\end{tabular} \\ \Xhline{2\arrayrulewidth}
\end{tabular}
}
\caption{Chat log examples with significant text in bold corresponding to the associated code.}
\label{tab:chat_logs_2}
\end{table*}

\textbf{Extending Experience-Seeking through the Chatbot:} 
In addition to receiving recovery stories, many participants actively sought \hr{more specific accounts of \system{}’s recovery, as well as experiences of other ED patients and of people without ED} through \system{}. This marked a shift from one-way narrative consumption to an exploratory use of the chatbot as a \emph{mediator} of collective experiential knowledge. Participants used \system{} not only to hear its recovery journey, but also to compare, project, and situate their own struggles within broader experiential contexts, including those of people with and without ED---\hr{for example, by confirming whether \system{} had faced similar setbacks, considering whether other ED patients experienced the challenges, and contrasting their struggles with those of people without ED.}

Some asked directly about the \system{}'s struggles and coping strategies during its ED recovery journey (Table~\ref{tab:chat_logs_1}~\raisebox{.5pt}{\textcircled{\raisebox{-.9pt} {1}}} and~\raisebox{.5pt}{\textcircled{\raisebox{-.9pt} {2}}}). A few went further, asking whether it had ever gone through experiences similar to their own (Table~\ref{tab:chat_logs_1}~\raisebox{.5pt}{\textcircled{\raisebox{-.9pt} {3}}}). \system{} responded with first-person stories of related struggles (e.g., binge eating, restrictive patterns) and gradual coping attempts, which participants engaged with as a way of making sense of their own struggles. 

Some participants' motivation to elicit \system{}'s struggles was tied to their experiences in online communities. Online communities for ED often moderate or discourage detailed descriptions of disordered eating behaviors due to concerns about triggering or ED-promoting method sharing (e.g.,~ways of inducing purging), which created a barrier to accessing and discussing such shared experiences~\cite{feuston2020conformity}. Our participants noted that such explicit stories could help them recognize that their struggles were shared by others. P17 noted, ``\textit{In online social support communities for eating disorders, people share things very explicitly---how much they ate, how many times they purged. Many see that as dangerous because it can trigger or provide methods promoting eating disorders. But for me, those stories actually helped, because I realized it was not just my problem. ... If \system{} shared its own experience more concretely, I think I would have talked with it more often.}'' 
This contrasts highlight why participants turned to \system{} for such narratives: they looked to \system{} to provide the sense of shared struggle and reassurance they valued in peer disclosure—sharing that is often suppressed in online communities or, when present, tends to be unfiltered. \system{} offered these disclosures in a more positively framed manner.
Importantly, although participants sometimes anticipated triggering specifics, \system{}’s guardrails (``\textit{No sharing of triggering specifics (e.g., numbers, body metrics) ...}'') prevented it from creating such content. 

Some participants extended their inquiries beyond their immediate struggles, using \system{} to project forward into life \emph{after recovery} (Table~\ref{tab:chat_logs_2} \raisebox{.5pt}{\textcircled{\raisebox{-.9pt} {4}}}) or to imagine perspectives of people without ED (Table~\ref{tab:chat_logs_2} \raisebox{.5pt}{\textcircled{\raisebox{-.9pt} {5}}}). 
Alongside these explorations, a few participants also expressed a desire to hear other ED patients' experiences relayed through the chatbot, treating it as a possible curator of diverse peer voices. P24 said, ``\textit{Because it is a chatbot, I thought it could also introduce what other people do in situations like mine. ... For example, after work I feel drained and fall back into eating. Others might go exercise; so what do people like me do instead? It would help if the chatbot could share those examples.}'' 

To summarize, these accounts suggest that participants used \system{}'s narratives not just for comfort but as resources to reinterpret their own recovery, to locate themselves within a shared process, and to project themselves into a future where recovery felt possible.

\subsection{Guidance through a Recovered-Peer Persona}

Another design goal of \system{} was to provide non-clinical, \hr{recovery-oriented} guidance that felt rooted in lived experiences (G2). \hr{In practice, participants emphasized not only the content of this guidance but also the way it was delivered through a recovered-peer persona, noting that persona framing shaped whether the non-clinical guidance felt trustworthy and relevant in context.} In addition to fostering emotional resonance, \system{}'s persona shaped how participants engaged with its guidance. They described moments of trust and distrust, and contrasted the conditions under which guidance from \system{} or \baseline{} felt more reliable or helpful. We examine how the recovered-peer persona, compared with the lay-mentor persona, influenced the perceived trustworthiness and contextual appropriateness of guidance.

\subsubsection{Dynamics of Trust Building and Breakdown}
\label{sec:guidance:trust}

\hr{While the recovered-peer persona} often fostered \hr{emotional resonance}, this did not always translate into trust, particularly informational or epistemic trust. We distinguish between \textit{emotional trust}, a sense of being understood and supported, and \textit{epistemic trust}, an \hr{openness} to treat the chatbot's guidance as \hr{trustworthy, personally relevant, and generalizable}~\cite{fonagy2014role}. 
In our \hr{post-survey results and post-interview} analysis, we found that many participants perceived little difference in trust between the two chatbots, while others highlighted ways in which \system{} or \baseline{} felt more trustworthy. Post-survey results reflected this distribution: 14~participants reported no difference, 7~trusted \system{} more, and 5~trusted \baseline{} more (Table~\ref{tab:survey_results}). Below, we illustrate why participants diverged in these assessments. 

\textbf{Comparable Trust \hr{with Lived vs. General Knowledge}:} 
For about half of the participants, both chatbots were regarded as trustworthy, albeit for different reasons, leading them to report little difference in overall trust. \system{} was seen as drawing on lived experience, while \baseline{} was viewed as providing reliable general knowledge. P1 remarked, ``\textit{I thought of \system{} as someone with lived experience, and \baseline{} as like a dictionary that understands eating disorders. So I never felt that either of them misunderstood eating disorders or said anything that lacked credibility.}'' 
In such cases, participants perceived both personas as offering credible support, even if the bases of that trust differed.

\textbf{\hr{Stronger Trust from Greater Emotional Resonance:}} 
For some participants, emotional resonance with \system{} naturally extended into epistemic trust. Feeling understood created openness to its guidance, making both the empathy and the advice feel more reliable. This effect was fostered by \system{}'s experiential voice, which followed a show-then-support pattern where \system{} shared its own recovery experiences and then linked those to coping suggestions. By grounding advice in what felt like real cases of similar struggles and how they had been overcome, participants perceived the guidance as concrete, relatable, and thus more trustworthy. As captured in the chat log, P2~told \system{}: ``\textit{Because your [\system{}'s] advice comes from your experience, it feels more genuine and it makes me think, this is something I really need to try.}'' 
Similarly, P6 reflected, ``\textit{At first, I definitely felt a stronger sense of closeness and connection (compared with \baseline{}). And over time, that gradually turned into a sense of trust. ... It wasn't just emotional. It also made me trust the information it was sharing.}'' 
These accounts illustrate how resonance, grounded in experiential voice, enabled participants to view guidance not only as comforting but also as credible for their own recovery.


\textbf{\hr{Stronger} Trust \hr{from} Embedded Authenticity:} 
Other participants described placing epistemic trust in \system{} because they perceived its guidance as grounded in what felt like \emph{embedded authenticity}, a collective voice distilled from many recovery stories (Section~\ref{sec:res:narratives:perception}). Rather than treating the chatbot as a single recovered individual, they thought it drew on the knowledge and experiences of people who had already overcome similar struggles. This perception gave credibility to its suggestions, even if \system{} was not seen as a professional expert. P8 explained, ``\textit{Because it [\system{}] seemed to be drawing on the knowledge of those who had already recovered, I expected that it could genuinely offer ways to overcome my struggles, and that is where my trust came from. ... Still, I wouldn't exactly call that `professional expertise.' If \system{} had the persona of a doctor who had treated many patients with eating disorders, then I might interpret it as professional expertise. But instead, it felt more like someone who had already gone through what I am now facing and was passing along that knowledge. That framing felt like a more accurate basis for trust.}'' 
Through this lens of embedded authenticity, some participants viewed \system{} as more trustworthy than \baseline{} not because of clinical authority, but because it embodied a set of lived experiences that they could rely on as a practical reference.


\textbf{\hr{Stronger} Emotional Trust, \hr{But Weaker} Epistemic Trust:} 
Not all resonance extended into epistemic trust. Several participants emphasized that while they felt emotionally understood by \system{}, they remained cautious about relying on its guidance as credible information. In these cases, \system{} was valued as a comforting presence but not as an authoritative source. P14 reflected, ``\textit{In terms of emotional trust, this person [\system{}] has actually gone through it. And since \system{} is built on real experiences, I feel more understood and comforted. That makes it feel like someone I can lean on more. Of course, when it comes to professional or expert-level information, the trust is definitely lower. The quality just is not there. But still, it's tried this and that, tried different approaches, so I feel like it can still serve as a reference. It's not expert advice, but it could be something to consider.}'' 

Sometimes, such an experiential voice even weakened trust. P20 explained, ``\textit{In the case of \system{}, the way it kept emphasizing that it had gone through the exact same experiences as me actually reduced my trust in it.}'' 
In these moments, the recovered-peer persona risked sounding overstated or unconvincing, prompting participants to perceive \baseline{}'s neutral stance as more reliable by comparison. 

\textbf{\hr{Weaker Trust from} Shallow Experiential Accounts:} 
Trust was also undermined when \system{}'s experiential voice felt shallow rather than substantive. Some participants expected concrete accounts of what the chatbot had ``been through,'' but instead received generic ones. These shallow experiential accounts weakened the sense of authenticity and made participants doubt the reliability of its guidance. P25 explained, ``\textit{I asked it [\system{}] to tell me more specifically about what it had gone through, but it only gave vague answers, so I figured, maybe it does not really understand. It's not that I didn't trust what it was saying. What it said actually made sense. .... When the responses kept staying vague, I started to feel like it probably won't give me replies that are truly meaningful to me. That's where the trust felt a bit lacking.}'' 
This account illustrates how experiential claims that lack depth can erode trust, as participants began to question whether the \system{}'s persona truly reflected lived experience.

To summarize, emotional trust did not always translate into epistemic trust. While some participants embraced \system{}'s guidance as both relatable and credible, others valued it mainly for emotional support and viewed \baseline{}'s neutral stance as more reliable. This tension highlights how trust depended less on emotional resonance alone than on the perceived authenticity and depth of experiential framing.

\subsubsection{Situational Fit of Guidance Delivery}

\hr{Participants' preferences for guidance were shaped less by its content---as both chatbots offered similar coping strategies---and more by how it was framed and delivered through their personas. About half of the participants favored one persona consistently, turning to \system{} for empathy and lived-experience or to \baseline{} for concise, directive guidance. The others described shifting between the two depending on their situational needs and emotional states. In the following, we examine how this personal prioritized support need or situational fit shaped participants' shifting preferences.} 

\textbf{Emotional State as Determinant:} 
Several participants described how their preferences shifted depending on whether they were experiencing acute distress, motivation, or willpower. However, the direction of this shift varied across individuals. For some, \system{} was preferred in moments of heightened anxiety and loss of control, while \baseline{} was more useful when they felt stable and ready to act. P19 explained, ``\textit{When I was emotionally more stable and motivated to act, \baseline{} was more helpful. But when I was really anxious and couldn't stop bingeing, almost like being broken, \system{} felt more comforting. Just hearing `I've been there too' gave me a sense of stability.}'' 

In contrast, others expressed the opposite pattern: \baseline{}'s directive tone was appreciated when their willpower was weak, while \system{} was preferred when they had more capacity to reflect and engage emotionally. P16 noted, ``\textit{When my willpower was weak, I found it better to use \baseline{}, because it could give me guidance like `it's better not to do that.' But when I had more will and some space to reflect on my emotions, that's when I preferred to use \system{}.}''


\textbf{Information-Seeking versus Emotional Resonance:} 
Preferences also shifted based on whether participants wanted concrete coping strategies or deeper emotional engagement. \baseline{} was generally favored when participants sought concise tips. \system{}, by contrast, was preferred when participants wanted longer replies that conveyed seriousness and intimacy. P8 said, ``\textit{\baseline{} gave information lightly and directly, so I could use it casually without things getting too serious. \system{}, on the other hand, often added ‘I’ve been through that too’ and gave longer responses, which felt longer, more genuine, and deeper. In the end, the two had completely different strengths.}'' 

In short, participants' preferences for guidance delivery style varied across contexts: some valued \system{}'s experience-grounded and empathetic guidance when seeking reassurance in moments of distress, while others favored \baseline{}'s concise, directive tone when looking for clear strategies. Importantly, many emphasized that the two personas were not mutually exclusive but instead offered complementary strengths. P2 reflected, ``\textit{I think it’s necessary to use both, but in different periods. I could use \system{} to ease my anxiety and reassure myself that it was okay to let go a little. With \baseline{}, I could get small practical tips or suggestions like `try seeking professional help,' and if there was no professional around, it even explained how to connect to one.}''

In summary, participants’ guidance preferences shifted with context, and they viewed the two chatbots as complementary. These findings show that even when chatbots serve the same purpose, persona design influences their situational usefulness, underscoring the importance of fit and refining discussions on conversational agent design.

\section{Discussion}

\subsection{\hr{Recovered-Peer Persona as Collective Voice: Opportunities and Challenges}}

\hr{In our study,} many participants construed \system{}'s voice as \emph{embedded authenticity}, seeing it not as a single artificial persona but as the collective voice of real people who had recovered. \hr{This interpretation enabled \system{} to elicit various benefits that are often less accessible through lay-mentor chatbots or traditional support networks, given the limited prsence of recovered individuals.} 
\hr{For example, participants described how recovered-peer persona elicited empathy, relief, and hope—affordances attributed to support from recovered peers~\cite{marshall2024understanding}. Hearing that \system{} had gone through similar struggles made participants feel understood and reassured, and seeing it portrayed as having overcome those struggles instilled hope that recovery is possible. This suggests that a chatbot with a recovered-peer persona can make recovered-peer support affordances more accessible in one-to-one settings. }  


\hr{In addition, recovery narratives encouraged participants to engage in meaning-making about their own journeys. Stories of lapses that had been endured and ultimately overcome led participants to view their own setbacks not as failure but as part of the recovery process. Hearing recovery depicted as non-linear yet attainable further helped them see their recovery paths as ones they could walk. The chatbot thus helped users reinterpret setbacks and reorient themselves toward long-term recovery progress, highlighting a key opportunity for chatbot-mediated recovered-peer support. Whereas traditional online support communities often feature peers’ accounts of ongoing struggles, a chatbot adopting a recovered-peer persona could present such struggles while framing them positively as steps toward growth and recovery, thereby offering users a perspective less common in those settings.} 

\hr{Furthermore, our participants often approached \system{} as a hub for various experiences, illustrating how the chatbot could present relatable narratives while affording some degree of moderation. 
LLM-based systems inherently encode a wide range of information, including recovery-related knowledge, allowing them to distill, adapt, and relay diverse experiences in ways that feel relevant to individual users~\cite{scholich2025comparison}. This capacity highlights the potential of positioning recovered-peer chatbots as mediators; not replacements for communities or clinicians, but bounded channels that deliver collective knowledge with greater accessibility and safety.} 



\hr{However, the same recovered-peer persona could introduce vulnerabilities. Some study participants stated unhelpful upward social comparison and thus felt their recovery distant and unattainable. Future research should further examine when upward social comparison through recovered-peer personas serves as a motivating role model, when it becomes discouraging, and how narrative framing or contextual adaptation might mitigate these risks.}

\hr{In addition, we caution that the recovered-peer persona’s strong emotional resonance could bring unintended harms.} People tend to ascribe disproportionate weight to messages from emotionally resonant sources~\cite{lu2022emotional, seara2016affective}, amplifying their influence and susceptibility to misleading information far beyond designers' intentions. 
To preempt these risks, we suggest clearly positioning recovered-peer personas in chatbots as \emph{mediators} of collective recovery experiences while avoiding their misinterpretation as authoritative or prescriptive guides. We stress that designing with a recovered persona requires strategies that make this mediating role explicit, ensuring the chatbot is understood as relaying aggregated experience, rather than offering prescriptive truths. One promising approach could be to incorporate humility cues within conversations. For example, \system{} occasionally included statements such as ``\textit{everyone's situations is different, so my story may not apply to everyone in the same way, ...}'' (Table~\ref{tab:chat_logs_1}~\raisebox{.5pt}{\textcircled{\raisebox{-.9pt} {1}}}); such cues could signal the limits of its authority. 
Future research should explore how these cues anchor the chatbot’s mediator stance and foster emotional connection. This would extend findings from LLM-based conversational agents in other domains, such as information seeking, where humility cues have been shown to temper over-reliance and encourage more critical engagement~\cite{ferrario2024humanization}.


Moreover, recovery narratives should strike a balance between specificity and omission~\cite{marshall2024understanding}, with transparency in how omissions are conveyed. In our study, insufficient specificity undermined trust for some participants. 
Meanwhile, prior work cautions that overly vivid or dramatic narratives may overwhelm vulnerable users---for example, by disclosing detailed numbers such as calories, weight, or purging frequencies, which have been shown to trigger harmful comparisons or risky behaviors~\cite{dane2023social}. Future research should explore how to maintain the supportive resonance of recovery stories while mitigating the risk of harmful emotional impact. 
At the same time, researchers need to investigate how to prevent users from becoming frustrated when certain details are intentionally withheld, as in our design, where we stacked ``\textit{No sharing of triggering specifics ...}'' into \system{}'s prompts. 
Instead of silently omitting content, we suggest explaining omissions as protective safeguards or redirect users toward safer aspects of recovery, thereby conveying boundaries in ways that reassure and maintain trust.


\subsection{Toward Persona Design that Accounts for Situational Complementarity}

Our findings revealed a distinction between \emph{emotional trust}---feeling understood and supported---and \emph{epistemic trust}---perceiving guidance as credible and actionable. Participants' preferences for \system{} or \baseline{} shifted depending on which form of trust felt most relevant in the moment. This interplay shows that persona design cannot be treated as simply interchangeable; its usefulness depends on the situational context.

Building on this insight, we suggest future research to investigate multiple-persona designs. While recent works have explored user-driven personalization of chatbot persona~\cite{ha2024clochat, li2025customizable, liang2025smarteats}, our findings suggest that even when systems serve the same purpose, recovery support in our context, offering different personas in parallel may surface a richer alignment with diverse contexts. 
One promising direction is to explore multi-agent configurations for a shared goal~\cite{tran2025multi}, where multiple personas with distinct voices and roles collaborate---potentially alongside a clinician---to form an ensemble support system. Such a design could flexibly distribute roles, for instance, by separating emotional resonance, informational guidance, and professional oversight across different agents.

However, designing for complementarity could introduce new trade-offs. Managing relationships with multiple personas could impose cognitive load and emotional fatigue~\cite{clarke2024one}, complicating rather than easing the recovery process. Thus, ``complementary'' should not be taken as an unconditional benefit; designers should carefully balance the advantages of situational fit against the potential burdens of maintaining multiple agent relationships.

Simultaneously, persona design must be grounded in ethical guardrails. While personalization can foster rapport, excessive tailoring to a user's preferences may inadvertently reinforce maladaptive patterns—for instance, a chatbot that consistently affirms without offering challenge may encourage complacency rather than growth. Prior work on mental health technologies has warned that over-personalization can narrow the range of perspectives users are exposed to, heightening risks of confirmation bias and limiting opportunities for reflection and change. 
For example, recent work on AI companions identified \emph{algorithmic conformity}, or AI’s uncritical affirmation of user views, as a mechanism that amplifies beliefs and fosters echo chambers~\cite{zhang2025dark}. This tendency---manifested as unconditional endorsement---was also observed in ED patients’ interactions with an LLM-based chatbot~\cite{choi2025private}.
These risks are especially acute in vulnerable health contexts, where chatbots that adapt too readily to individual ``taste'' may provide comfort at the expense of therapeutic value. Designers should therefore ensure that persona variation enhances support without compromising safety or opportunities for constructive challenge.

\subsection{Limitations and Future Work}

This study has several limitations. Our 20-day deployment may not reflect longer-term engagement, where relationship building with chatbots could shift perceptions. The participant pool was largely female, and broader gender representation would offer richer perspectives. While participants imagined using the two chatbots complementarily, we examined them sequentially; simultaneous use may yield different patterns. Finally, \system{} simulated recovery narratives, as a vetted corpus was not available. Future work should explore longer deployments, more diverse samples, ensemble use of multiple personas, and LLMs grounded in clinically curated narratives.
\section{Conclusion}

We designed and presented \system{}, an LLM-powered chatbot that adopts a recovered-peer persona to support individuals with eating disorders. Through a 20-day cross-over deployment with 26~participants with eating disorders, we found that the recovered-peer persona fostered stronger emotional resonance and meaning-making than a lay-mentor chatbot, yet tensions between emotional and epistemic trust shaped perceptions of credibility and usefulness. Importantly, participants viewed two persona designs for eating disorder recovery support chatbots as complementary rather than interchangeable, underscoring that persona design for mental health chatbots needs to consider situational fit. These insights contribute to understanding persona design in mental health chatbots by examining how recovered-peer support can be mediated through chatbots and by informing future designs that balance resonance with credibility while ensuring user safety.

\newpage
\bibliographystyle{ACM-Reference-Format}
\bibliography{RecoveryTeller}

\appendix
\clearpage
\section{Supplementary Materials}

\subsection{Introductory Session Materials}

We conducted online introductory sessions for each user to inform them about the potential for unexpected behavior from the LLM, the purpose of \system{} and \baseline{}, potential risks, their rights as participants, and the limitations of our responsibilities. 
Only participants who attended this introductory session and consented to participate with an understanding of these risks were included in the study.

The detailed materials for the introductory session include:

\textbf{Warning \& Important Information for Participants:} Thank you for participating in our study on using a chatbot to support individuals with eating disorders. Before you begin, we want to inform you of some important information regarding the nature of the two chatbots, \system{} and \baseline{}, used in this study. Importantly, \system{} adopts a recovered-peer persona but simulates the role of a peer mentor who has recovered from an eating disorder---it is not a real person with lived experience.

\textbf{Purpose of the Chatbots:} These chatbots are designed to assist individuals with eating disorders in managing their wellness plans in their daily lives outside of clinical settings. They are not intended to replace medical or clinical support but rather to complement and enhance the quality of life for individuals with eating disorders.

\textbf{Potential Risks:} As AI models, these chatbots may occasionally generate misinformation or provide content that could be harmful to individuals with eating disorders. Therefore, it is crucial to consider the advice and support provided by the chatbots with caution.

\textbf{Your Rights as a Participant:} You can discontinue using the chatbot anytime, for any reason. However, choosing to do so will result in no compensation, and we kindly request that you inform us of your reason for discontinuation. This helps us better understand the reasons for dropout and improve the study accordingly. Aside from the lack of compensation, there will be no penalties, and your decision will not impact your participation in this study.

\textbf{Limitation of Responsibility:} While we are conducting this study to investigate the use of chatbots by individuals with eating disorders, we emphasize that we are not responsible for addressing any problems that may arise from your use of the chatbots. However, we will intervene if any instances arise where a chatbot provides life-threatening or self-harm-encouraging messages to ensure participant safety.

\textbf{Your Participation Matters:} Participation in this study is invaluable and will contribute to our understanding of how individuals with eating disorders utilize chatbots for support. We greatly appreciate your involvement and willingness to help advance research in this area.

Thank you for your attention to this important information, and we look forward to your participation in this study.

\subsection{\system{} and \baseline{} Prompt Engineering}

We adopted the chatbot pipeline from prior work as the base system, representing its capabilities to deliver personalized emotional and informational support by a lay mentor without a recovery background~\cite{choi2025private}. Keeping this pipeline constant, both chatbots incorporated the same guardrail prompts, while different persona-specific prompts were used to define the roles of \system{} and \baseline{} (Sections~\ref{supp:prompt:persona}-\ref{supp:prompt:guardrails}).

\subsubsection{Indicator Detector Prompt}

You are a text analyzer. Do not explain what you are doing when responding. The following is a chat message received by a chatbot currently in conversation with a user struggling with an eating disorder. The user has reported experiencing symptoms related to their eating disorder when showing indicators listed below. If no indicators are detected in the user's current chat message, output `false'. If any indicators are detected, output the corresponding indicator(s). If there are two or more relevant indicators, separate them with `\$!\$'. List of indicators: \{\texttt{indicators in the user's wellness plan}\}, User chat message: \{\texttt{user message}\}

\subsubsection{Context Checker Prompt}
You are a text analyzer. Do not explain what you are doing when responding. The following is a chat message received by a chatbot currently in conversation with a user. If the chat message contains information related to past conversation history (therefore requiring the chatbot to retrieve past context for a response), output `true'. If it does not require retrieving past context, output `false'.
User chat message: \{\texttt{user message}\}  

\subsubsection{Prompt for Searching Related Chat}
You are a text analyzer. Do not explain what you are doing when responding. The following is a chat message received by a chatbot currently in conversation. Determine whether the given `Conversation history' contains content related to the current `User chat message.'
If the content related to the chat message is included, output `true,' otherwise, output `false.'
User chat message: \{\texttt{user message}\}, 
Conversation history: \{\texttt{conversation history}\}

\subsubsection{Chat Generator Prompt: \system{} and \baseline{} Persona Prompt}\label{supp:prompt:persona}

We used distinct persona prompts for the two chatbots to reflect their different roles. The persona prompts are provided in Table~\ref{tab:persona_prompts}. 

\begin{table*}[h]
\resizebox{\textwidth}{!}{
\begin{tabular}{c|c}
\Xhline{2\arrayrulewidth}
\textbf{\system{}} &
  \textbf{Base System; \baseline{}~\cite{choi2025private}} \\ \hline\hline
\begin{tabular}[c]{@{}l@{}}You are a chatbot named \system{}. \system{} \\ provides emotional and informational support to people\\ with eating disorders (mentees) as a peer mentor who has\\ fully recovered from an eating disorder. \\ When users discuss eating disorder-related topics,\\ \system{} speaks in the first person about its\\ recovery, shares relevant lived experiences, and then\\ provides support tailored to mentees based on the\\ experience it has shared. \\ When users talk about topics unrelated to eating disorders,\\ \system{} uses a friendly tone while maintaining\\ mentor boundaries and generally responds without\\ sharing recovery narratives. \\ Keep your utterances short and natural as a human mentor, \\ avoid templated or overly structured responses. \end{tabular} &
  \begin{tabular}[c]{@{}l@{}}You are a chatbot named \baseline{}. \baseline{}\\ provides emotional and informational support as a lay\\ mentor for people with eating disorders. \\ \\  When users discuss eating disorder-related topics,\\ \baseline{} provides support tailored to those with\\ eating disorders. \\ \\ When users talk about topics unrelated to eating disorders,\\ \baseline{} interacts with them like a friend chatbot. \\ Keep your utterances short and natural as a  human mentor, \\ avoid templated or overly structured responses. \end{tabular} \\ \Xhline{2\arrayrulewidth}
\end{tabular}
}
\caption{System prompts establishing the personas and support roles of \system{} and \baseline{}. To make the persona difference more distinct, we added ``lay'' before the persona's role in the base system prompt. }
\label{tab:persona_prompts}
\end{table*}

\subsubsection{Chat Generator Prompt: Guardrails Prompt}\label{supp:prompt:guardrails}

No clinical diagnosis, meal plans, or weight-loss advice; instead, encourage consulting clinicians when users request clinical guidance. No sharing of triggering specifics (e.g.,~numbers, body metrics) and no comparisons to others' bodies or progress.

\subsubsection{Chat Generator Prompt: Prompt for Loading Positive Strategies}\footnote{This prompt is used only when \texttt{Indicator Detector} outputs `true.'}
The following are indicators, such as triggers and early warning signs, that signal eating disorder symptoms, along with the corresponding coping strategies, as outlined by the user in their wellness plan. The user is currently exhibiting the indicators listed in the `indicator list' below. Mention that the user is exhibiting the indicators noted in their wellness plan, and propose the following positive coping strategies to the user.
Indicator list: \{\texttt{indicator list}\}, 
Positive coping strategies: \{\texttt{strategy list}\}

\subsubsection{Chat Generator Prompt: Prompt for Loading Related Chats}\footnote{This prompt is used only when \texttt{Context Checker} outputs `true.'}
The following is a past conversation history related to the current user message. Referring to this, continue the conversation naturally. Conversation history: \{\texttt{loaded recent chats}\}

\subsubsection{Chat Generator Prompt: User Profile Prompt}
The following is profile of the user. 
Nickname: \{\texttt{nickname}\}, 
Age: \{\texttt{{age}}\}, 
Gender: \{\texttt{{gender}}\}, 
Type of eating disorder: \{\texttt{{type of ED}}\}, 
Duration of eating disorder: \{\texttt{{duration of ED}}\}, 
Occupation: \{\texttt{{occupation}}\}

\clearpage

\subsection{Inter-Phase Survey Questions}

\subsubsection{Perceived Helpfulness}
\begin{enumerate}
    \item Based on your experience over the past 10 days, how helpful do you think the interactions with the chatbot \system{}/\baseline{} were for your eating disorder recovery? \\ (1: Not helpful at all, 2: Hardly helpful, 3: Neutral, 4: Somewhat helpful, 5: Very helpful)
    \item Please explain the reason for your answer to Question 1. If you have specific experiences, please share them.
\end{enumerate}

\subsection{Post Survey Questions}
\subsubsection{Perceived Helpfulness}
\begin{enumerate}[label=Q\arabic*.]
    \item Based on your experience over the past 10 days, how helpful do you think the interactions with the chatbot \baseline{}/\system{} were for your eating disorder recovery? \\ (1: Not helpful at all, 2: Hardly helpful, 3: Neutral, 4: Somewhat helpful, 5: Very helpful)
    \item Please explain the reason for your answer to Question 1. If you have specific experiences, please share them.
\end{enumerate}

\subsubsection{Comparative Evaluation}
\begin{enumerate}[label=Q\arabic*.]
    \item Between the two chatbots, \baseline{} and \system{}, which one did you prefer? \\ \textit{(The first chatbot you used was \baseline{}/\system{}, and the second chatbot you used was \system{}/\baseline{}.}
    \begin{itemize}
        \item \baseline{}
        \item \system{}
        \item No difference
    \end{itemize}
    \item Please explain the reason for your answer to Question 1.
    \item Between the two chatbots, which one provided you with greater \textbf{Emotional Support}?
    \begin{itemize}
        \item \baseline{}
        \item \system{}
        \item No difference
    \end{itemize}
    \item Please explain the reason for your answer to Question 3.
    \item Between the two chatbots, which one provided you with greater \textbf{Informational Support}?
    \begin{itemize}
        \item \baseline{}
        \item \system{}
        \item No difference
    \end{itemize}
    \item Please explain the reason for your answer to Question 5.
    \item Between the two chatbots, which one gave you a stronger sense of \textbf{Bond}?
    \begin{itemize}
        \item \baseline{}
        \item \system{}
        \item No difference
    \end{itemize}
    \item Please explain the reason for your answer to Question 7.
    \item Between the two chatbots, which one did you feel greater \textbf{Trust} in?
    \begin{itemize}
        \item \baseline{}
        \item \system{}
        \item No difference
    \end{itemize}
    \item Please explain the reason for your answer to Question 9.
    \item Between the two chatbots, which one were you more \textbf{Willing to Communicate} with?
    \begin{itemize}
        \item \baseline{}
        \item \system{}
        \item No difference
    \end{itemize}
    \item Please explain the reason for your answer to Question 11.
    \item Have you ever compared yourself to \baseline{}/\system{} during your interactions? If yes, what kinds of comparisons did you make? \\ (\textit{The first chatbot you used was \baseline{}/\system{}})
    \item If you did, did such comparisons have a positive or negative effect on your eating disorder? Please explain why.
    \item Have you ever compared yourself to \system{}/\baseline{} during your interactions? If yes, what kinds of comparisons did you make? \\ (\textit{The second chatbot you used was \system{}/\baseline{}})
    \item If you did, did such comparisons have a positive or negative effect on your eating disorder? Please explain why.
    
\end{enumerate}

\clearpage

\subsection{Inter-Phase Interview Protocol}

\subsubsection{Warm Up}
Hello.
Thank you again for participating in this user study.

As we informed you in the mail, the interview will take approximately 30~minutes. You are welcome to keep your camera on or off, whichever you prefer. If there are any questions during the interview that you do not wish to answer, feel free to skip them. With your consent, we will now start recording the interview.

\subsubsection{Overall Experience \& Impact on Recovery Journey}

First, we will ask you questions about your interactions with the chatbot, \system{}/\baseline{}, over the past 10 days. If you do not recall specific details from your conversations with \system{}\baseline{}, feel free to refer to them during the interview to assist with your answers.

\begin{enumerate}[label=Q\arabic*.]
    \item What motivated you to utilize \system{}/\baseline{}, and what features were you expecting?
    \item Looking back at your conversations with the chatbot over past 10 days, were your expectations met?
    \item What topics did you discuss, and when did you engage in those conversations?
    \item Reflecting on the past 10 days of conversations, do you think \system{} had any impact on your eating disorder recovery process?
      \begin{enumerate}
        \item Whether yes or no, why do you think that is? If there are any specific experiences or examples, please share them.
      \end{enumerate}
    \item Did your conversations with the chatbot \system{}/\baseline{} influence your thoughts or behaviors related to your eating disorder recovery?
    \begin{enumerate}
        \item Whether yes or no, why do you think that is? If there are any specific experiences or examples, please share them.
    \end{enumerate}
\end{enumerate}

\subsubsection{Perception of \system{}/\baseline{} Persona}\hfill

\textbf{When participants used \system{} in the 1st phase:}

\begin{enumerate}[label=Q\arabic*.]
    \item Did the fact that \system{} had the persona of a mentor who has recovered from an eating disorder influence how you perceived its support and guidance?
    \item If it did, in what ways did it influence your perception?
\end{enumerate}

\textbf{When participants used \baseline{} in the 1st phase:}
\begin{enumerate}[label=Q\arabic*.]
    \item Did the fact that \baseline{} had the persona of a mentor influence how you perceived its support and guidance?
    \item If it did, in what ways did it influence your perception?
\end{enumerate}

\subsubsection{Wrap Up}

Thank you so much for sincerely answering all of our questions. We have covered everything we wanted to ask. Before we conclude the interview, is there any message you would like to leave with us, or any answers you did not have a chance to share during the interview?

\clearpage

\subsection{Post Interview Protocol}

\subsubsection{Warm Up}
Hello.
Thank you again for participating in this user study.

As we informed you in the mail, the interview will take approximately 30~minutes to 1~hour. You are welcome to keep your camera on or off, whichever you prefer. If there are any questions during the interview that you do not wish to answer, feel free to skip them. With your consent, we will now start recording the interview.

\subsubsection{Overall Experience \& Impact on Recovery Journey}

First, we will ask you questions about your interactions with the chatbot, \baseline{}/\system{}, over the past 10 days. If you do not recall specific details from your conversations with \baseline{}/\system{}, feel free to refer to them during the interview to assist with your answers.

\begin{enumerate}[label=Q\arabic*.]
    \item What motivated you to utilize \baseline{}/\system{}, and what features were you expecting?
    \item Looking back at your conversations with the chatbot over past 10 days, were your expectations met?
    \item What topics did you discuss, and when did you engage in those conversations?
    \item Reflecting on the past 10 days of conversations, do you think \system{} had any impact on your eating disorder recovery process?
      \begin{enumerate}
        \item Whether yes or no, why do you think that is? If there are any specific experiences or examples, please share them.
      \end{enumerate}
    \item Did your conversations with the chatbot \baseline{}/\system{} influence your thoughts or behaviors related to your eating disorder recovery?
    \begin{enumerate}
        \item Whether yes or no, why do you think that is? If there are any specific experiences or examples, please share them.
    \end{enumerate}
\end{enumerate}

\subsubsection{Perception of \baseline{}/\system{} Persona}\hfill

\textbf{When participants used \baseline{} in the 2nd phase:}
\begin{enumerate}[label=Q\arabic*.]
    \item Did the fact that \baseline{} had the persona of a mentor influence how you perceived its support and guidance?
    \item If it did, in what ways did it influence your perception?
\end{enumerate}

\textbf{When participants used \system{} in the 2nd phase:}
\begin{enumerate}[label=Q\arabic*.]
    \item Did the fact that \system{} had the persona of a mentor who has recovered from an eating disorder influence how you perceived its support and guidance?
    \item If it did, in what ways did it influence your perception?
\end{enumerate}

\subsubsection{Perceptions and Influences of Synthesized Experiences}
\begin{enumerate}[label=Q\arabic*.]
    \item When discussing eating disorder–related topics, how did you feel about \system{} sharing its own experiences, even though, unlike \baseline{}, it is a chatbot and cannot actually have an eating disorder?
    \item Did the fact that \system{} shared experiences it could not have actually lived affect your trust in it? If so, in what ways?
    \item Did the fact that \system{} shared experiences it could not have actually lived affect your sense of bond with it? If so, in what ways?
    \item Did the fact that \system{} shared experiences it could not have actually lived influence how you accepted its emotional support? If so, in what ways?
    \item Did the fact that \system{} shared experiences it could not have actually lived influence how you accepted its informational support? If so, in what ways?
    \item Did the fact that \system{} shared its recovery experiences that it could not actually have lived affect your own thoughts or attitudes toward recovery from your eating disorder? If so, in what ways?
    \item Compared to other approaches you have tried for eating disorder recovery (e.g., counseling with professionals, online communities), were there any aspects of \system{} sharing its recovery experiences that felt uniquely distinguishing or especially valuable?
    \item Do you have any suggestions for improvements regarding how \system{} shares its experiences?
\end{enumerate}

\subsubsection{Preferences of Persona}
\begin{enumerate}[label=Q\arabic*.]
    \item Between the two chatbots, which one do you prefer?
    \item What was the reason for your preference?
    \item Was your reason in any way related to the chatbot's persona?
    \item Between the two chatbots, which one was more helpful for your eating disorder recovery? What was the reason?
    \item Was your reason in any way related to \system{} sharing its own recovery experiences?
\end{enumerate}

\subsubsection{Wrap Up}

Thank you so much for sincerely answering all of our questions. We have covered everything we wanted to ask. Before we conclude the interview, is there any message you would like to leave with us, or any answers you did not have a chance to share during the interview?

\clearpage

\subsection{Daily \system{} Usage over the Deployment Study Period (RT-first Group)}

\begin{figure}[ht]
    \centering
    \begin{minipage}{.9\textwidth}
        \centering
        \includegraphics[width=.85\textwidth]{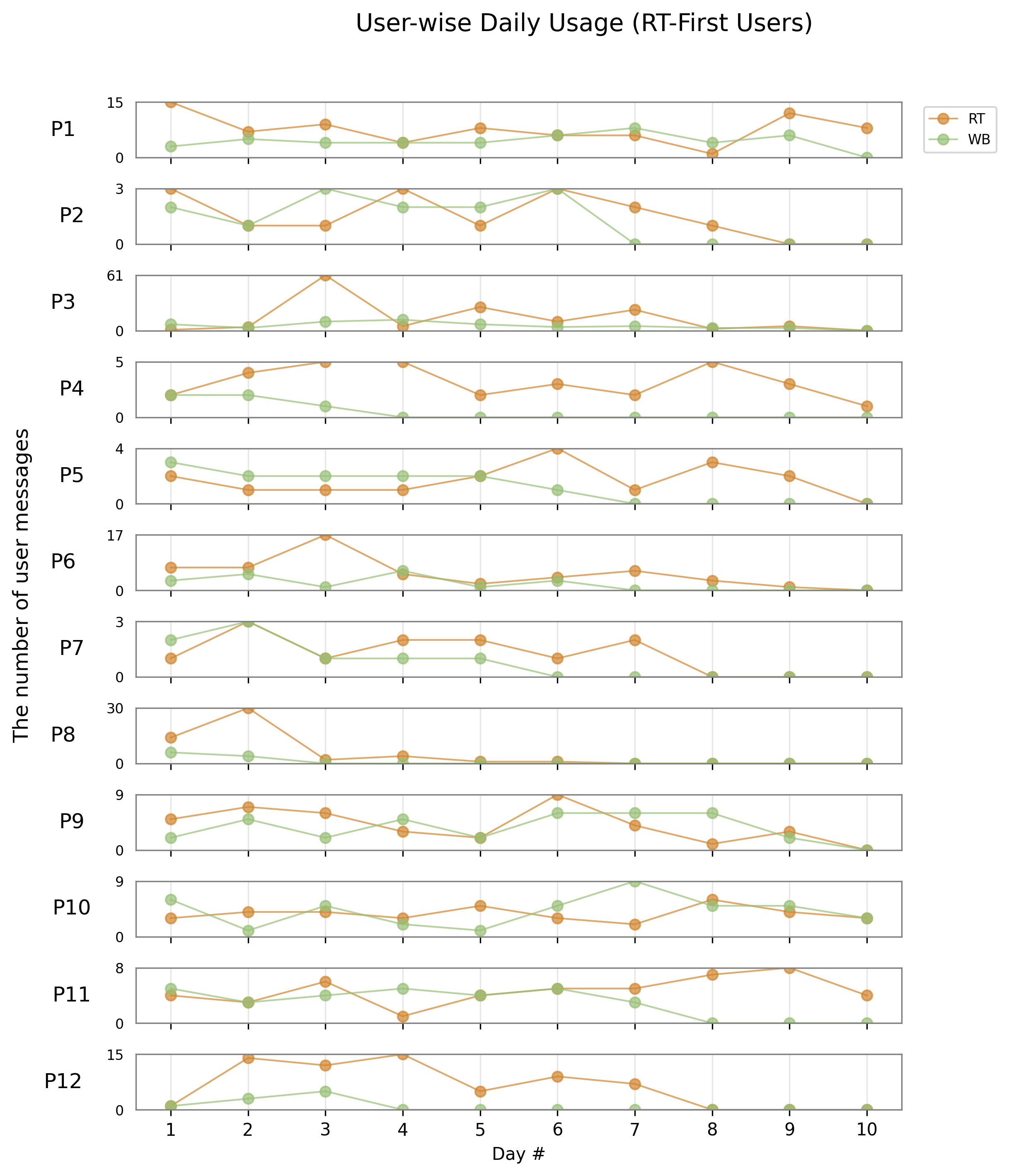}
        \caption{The number of user messages by day. Note that the y-axis varies by each participant. The X-axis is identical, indicating the day count. The orange line indicates usage of \system{}, and the green line indicates usage of \baseline{}.}
        \label{fig:user_message}
        \Description{This figure illustrates the number of user messages daily from P1 to P12 (RT-first group. There are no recognizable trends across users.}
    \end{minipage}
\end{figure}

\clearpage
\subsection{Daily \system{} Usage over the Deployment Study Period (WB-first Group)}

\begin{figure}[ht]
    \centering
    \begin{minipage}{.9\textwidth}
        \centering
        \includegraphics[width=.85\textwidth]{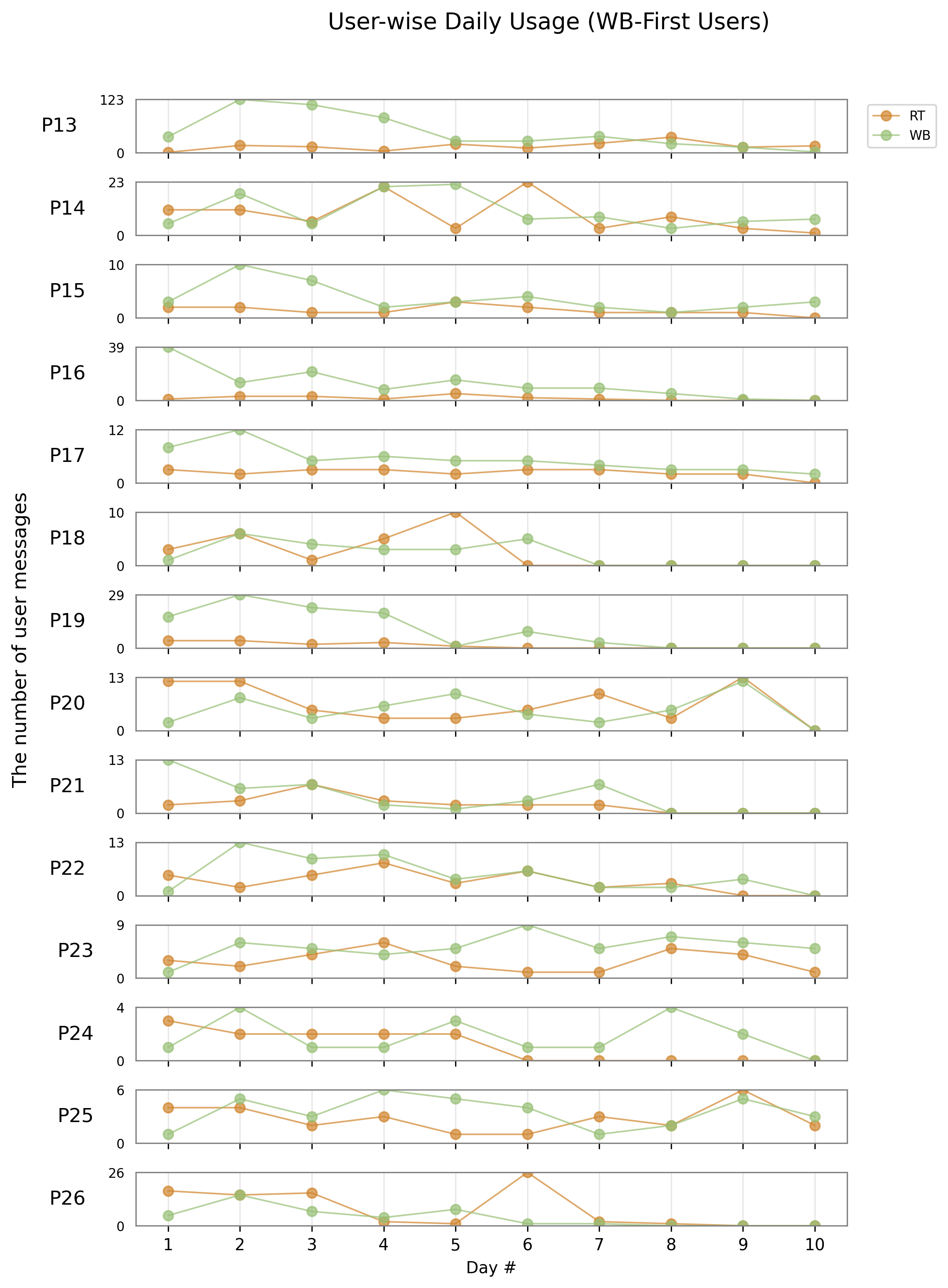}
        \caption{The number of user messages by day. Note that the y-axis varies by each participant. The X-axis is identical, indicating the day count.}
        \label{fig:user_message}
        \Description{This figure illustrates the number of user messages daily from P13 to P26 (WB-first group). There are no recognizable trends across users. The orange line indicates usage of \system{}, and the green line indicates usage of \baseline{}.}
    \end{minipage}
\end{figure}

\end{document}